\documentstyle[twoside,aps,prl,twocolumn,epsf,floats]{revtex}
\begin{document}
\textwidth=16truecm
\title{\underline 
{\rm PHYSICAL REVIEW E ~~~~~~~~~~~~~~~~~~~~~~~~~~~~~~~~~
~~~~~~~~~~~~~~~~~~~~~~~~~~~~~~~~~~~~~~
          in press}\\~~\\
An Optimal Shell Model of Turbulence} 
\author {Victor S. L'vov$^{*}$,  Evgenii Podivilov$^{\dag *}$,  
Anna Pomyalov$^*$,  Itamar Procaccia$^*$
and Damien Vandembroucq$^*$ \\
$^*$Department of~~Chemical Physics,
  The Weizmann Institute of Science, Rehovot 76100, Israel\\
  $^{\dag}$Institute of Automatization, 
  Russian Academy of Science,
  Novosibirsk 630090 , Russia.}  
\maketitle
\begin{abstract}
  We introduce a new shell model of turbulence which exhibits improved
  properties in comparison to the standard (and very popular) GOY
  model. The nonlinear coupling is chosen to minimize correlations
  between different shells. In particular the second order correlation
  function is diagonal in the shell index, the third order correlation
  exists only between three consecutive shells. Spurious oscillations
  in the scaling regime, which are an annoying feature of the GOY
  model, are eliminated by our choice of nonlinear coupling.  We
  demonstrate that the model exhibits multi-scaling similarly to these
  GOY model. The scaling exponents are shown to be independent of the
  viscous mechanism as is expected for Navier-Stokes turbulence and
  other shell models. These properties of the new model make it
  optimal for further attempts to achieve understanding of
  multi-scaling in nonlinear dynamics.
\end{abstract}
\section{Introduction}
Shell models of turbulence
\cite{Gledzer,GOY,Jensen91PRA,Piss93PFA,Benzi93PHD} are
simplified caricatures of the equations of fluid mechanics in
wave-vector representation; typically they exhibit anomalous scaling
even though their nonlinear interactions are local in wavenumber
space. Their main advantage is that they can be studied via fast and
accurate numerical simulations, in which the values of the scaling
exponents can be determined very precisely. Our interest in shell
models stemmed from our efforts to develop analytic methods for the
calculation of the numerical values of the scaling
exponents\cite{98BLPP}. In trying to do so we discovered that the most
popular shell model that was treated in the literature, the so-called
GOY models \cite{Gledzer,GOY}, poses very tedious calculations because
it exhibits slowly decaying correlations between velocity components
with different wave-numbers. In addition it has large oscillations
around the power law behavior in the scaling regime, making the
numerical calculation of the scaling exponents less obvious than
advertised. We therefore derived a new model which exhibits similar
anomalies of the scaling exponents but much simpler correlation
properties, and much better scaling behavior in the inertial range.
Since there is a significant number of researchers that are interested
in this type of models independently of the analytic calculability of
the exponents, we decided to present the model {\em per se}, discuss
its good properties, display the results of numerical simulations, and
compare it to the standard GOY model. These are the aims of this
paper.

In Section 2 we review the popular GOY model, and explain the
shortcomings that induced us to consider a new model. Section 3
introduces the new model, that we propose to call the Sabra model; we
discuss the phase symmetries and correlations, stressing the much
improved properties. 
Section 4 discusses numerical simulations from the algorithmic point
of view. Section 5 contains the results of numerical simulations and
fit procedures for accurate calculations of the scaling exponents. We
believe that this section contains methods that should be used in the
context of any shell model, and go beyond naive log-log plots. Section
6  presents a discussion of the limitations in computing high order
exponents. We demonstrate that beyond $\zeta_8$ one needs
exponentially long running times to extract reliable exponents. The
evaluation of $\zeta_{10}$ requires about one million turn-over times
of the largest scales.  We believe that similar limitations are
important also in other examples of multiscaling, including
Navier-Stokes turbulence. Section 7 demonstrates the universality of
the scaling exponents with respect to the viscous mechanism, and
Section 8 offers a short summary.
\section{Review of the GOY model}
\subsection{Basic Properties}
In the past considerable attention has been given to one particular
version of shell models, the so-called GOY model \cite{Gledzer,GOY}.
This model describes the dynamics of a complex ``Fourier" component of
a scalar velocity field that is denoted as $u_n$.  The associated
wavenumber is 1-dimensional, denoted as $k_n$.  The index $n$ is
discrete, and is referred to as the ``shell index".  The equations of
motion read: 
\begin{eqnarray}\label{goy}  \frac{d u_n}{dt}&=& i\big( ak_{n+1}
u_{n+2}u_{n+1} + bk_n u_{n+1}u_{n-1}
\\ \nonumber 
&&+ ck_{n-1} u_{n-1}u_{n-2}
\big)^* -\nu k_n^2  u_n +f_n \,, \end{eqnarray} 
where the star stands for complex
conjugation. The wave numbers $k_n$ are chosen as a geometric
progression \begin{equation}\label{kn} k_n=k_0 \lambda^n\,, 
\end{equation} 
with $\lambda$ being the ``shell spacing'' parameter.  $f_n$ is a
forcing term which is restricted to the first shells.  The parameter
$\nu$ is the ``viscosity".  In the limit of zero viscosity one can
arrange the model to have two quadratic invariants. Requiring that the
energy
\begin{equation}\label{energy}
E=\sum_n |u_n|^2\,, 
\end{equation} 
will be conserved leads to the following
relation between the coefficients $a$, $b$ and $c$:
\begin{equation}\label{abc}
a+b+c=0\ .
\end{equation}
A second quadratic quantity that is conserved is then
\begin{equation}\label{Hinv}
H=\sum_n (a/c)^n|u_n|^2
\ .
\end{equation}

Although non positive, this second invariant is often associated with
``helicity''. 

The main attraction of this model is that it displays multiscaling in
the sense that moments of the velocity depend on $k_n$ as power laws
with nontrivial exponents:
\begin{equation}
\label{scaling}
\langle |u_n|^q\rangle \propto k_n^{-\zeta_q}\,,
\end{equation}
 where the scaling exponents $ \zeta_q$ exhibit non linear 
dependence on $q$. 
We expect such scaling laws to appear in the ``inertial range"
with shell index $n$ much larger than the largest shell index that
is effected by the forcing, denoted as $n_L$, and much smaller
than the shell indices affected by the viscosity, the smallest
of which will be denoted as $n_d$.

We will refer to the moments as ``structure functions".
For even $q=2m$ we use the usual definition:
\begin{equation}\label{S2m}
S_{2m}(k_n)=\langle |u_n|^{2m}\rangle\,, 
\end{equation}
while for odd $q=2m+1$ we suggest the following definition:
\begin{eqnarray}      \nonumber 
 S_{2m+1}(k_n)={\rm Im}\langle u_{n-1}u_n &u_{n+1}& |u_n|^{2(m-1) }
\rangle\,,\\ 
&&  {\rm (GOY)}\ .\label{S2m+1}
\end{eqnarray}
The definition of the odd structure function differs from the usual
definition $S_{2m+1}(k_n)=\langle |u_n|^{2m+1} \rangle$. Our choice
Eq.~(\ref{S2m+1}) is motivated by our reluctance to use the
nonanalytic function $|u_n|$.  We will see that our definition yields
$\zeta_3=1$ as an exact result, similar to Kolmogorov's exact result
for $\zeta_3$ in 3-dimensional fluid turbulence. It was shown by
numerical simulations that the choice of parameters $\lambda=2$ and
$(a,b,c)=(1,-0.5,-0.5)$ leads to scaling exponents $\zeta_q$ that are
numerically close to those measured in experimental hydrodynamic
turbulence.

\subsection{Additional Properties}
\label{ssect:2.2}
The GOY model shares with Navier-Stokes turbulence an
analog of the 4/5 law. Assuming stationarity and using the
quadratic invariants introduced above, we can obtain two identities
involving third order correlations. Multiplying Eq.~(\ref{goy}) 
by $u_n^*$  we have, neglecting viscosity:
\begin{eqnarray}\label{flux}
{d\over d t}S_2(k_n)&=&
2k_0\lambda^n \Big[a\lambda S_3(k_{n+1})+bS_3(k_n) \\ \nonumber
&& +{c\over \lambda} S_3(k_{n-1})\Big]+p_n\ .
\end{eqnarray}
where 
\begin{equation}
\label{pump}
p_n=2{\rm Re}\left< u_n^* f_n \right>\ ,
\end{equation}
and obviously $p_n=0$ for $n>n_L$. In stationary conditions the
rate of change of $S_2(k_n)$ vanishes, and we find
\begin{eqnarray}\label{flux1}
a\lambda S_3(k_{n+1})+bS_3(k_n)+{c\over \lambda}S_3(k_{n-1})=0 \ .
\end{eqnarray}
This equation has a solution in the inertial interval:
\begin{equation}\label{solS}
S_3(k_n)= {1\over k_n}\Big[A +B  \Big({c\over a}\Big)^n\Big]\ .
\end{equation}
The unknown coefficients $A$ and $B$ can be found by its matching with
the ``boundary conditions'' at small $k_n$.  To do so we can follow
the considerations of Pissarenko {\it et al} \cite{Piss93PFA} and 
sum up Eq.(\ref{flux}) on all the shells from $n=0$ to an arbitrary
shell $M$, where $M$ is in the inertial
interval.  Using the conservation laws (i.e. a+b+c=0) we derive
\begin{eqnarray} \label{cond1}
0&=&{d\over dt}\sum\limits_{n=0}^{M} S_2(k_n) \\ \nonumber
&=&2k_M\Big[ a\lambda S_3(k_{M+1})
+(b+a)S_3(k_M) \Big] + \overline\epsilon \,, \\
 \label{cond2}
0&=&{d\over dt}\sum\limits_{n=0}^{M} S_2(k_n) \Big({a\over c}\Big)^n 
\\ \nonumber 
 &=&2k_M\Big({a\over c}\Big)^M
\Big[ a\lambda S_3(k_{M+1})
+(b+c)S_3(k_M) \Big] + \overline\delta \,,
\end{eqnarray}
where the rate of dissipation $\overline\epsilon$ and 
the spurious mean $\overline\delta$ are defined as
\begin{equation}\label{eps-delta}
\overline\epsilon =\sum \limits_{n=0}^{n_L} p_n\,,\quad
\overline\delta=\sum \limits_{n=0}^{n_L} p_n\Big({a\over c}
\Big)^n\ .
\end{equation}
Substituting the solution (\ref{solS})  into Eqs.
(\ref{cond1}), (\ref{cond2}) one relates the values of $A$ and $B$ to 
those of the fluxes $\overline\epsilon$ and $\overline\delta $. Now Eq. (\ref{solS}) 
becomes
\begin{equation}\label{al}
S_3(k_n)= {1\over 2 k_n(a-c)}
\Big[-\overline\epsilon
+ \overline\delta
\Big({c\over a}\Big)^n\Big]\ .
\end{equation}
\begin{figure}
\epsfxsize=7.5truecm\epsfbox{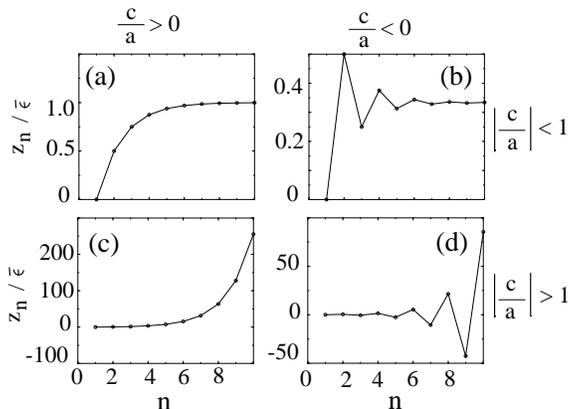}
\caption{Plots of the quantities $z_n =- k_n S_3(k_n)$ with $S_3(k_n)$
taken as the stationary solutions (16). The fluxes  
$\overline\epsilon$ and $\bar \delta$ are related by $\bar \epsilon = 
c\overline\delta/a$ and  $a=1$. The four panels have different values
of $c$. Panel  (a): $c=0.5$, Panel (b): $c=-0.5$, Panel (c): $c=2$, Panel (d):
  $c=-2$}
\vskip .1cm
\label{Fig1} 
\end{figure}
There are four different types of functional dependence of $S_3(k_n)$ on
$k_n$, determined by
the ratio $c/a$, as illustrated at Fig~\ref{Fig1}.  For $c/a<0$ this
function has period-two oscillations which are caused by the existence of
a non-zero flux
of the second integral of motion which is not
positively defined in this region.  For $c/a>0 $ the second
integral is positively defined and the function is monotonic.  
For $|c/a| < 1 $ the role of the second flux becomes
irrelevant in the limit $n\to\infty$. Consequently the deviation of $S_3(k_n)$
from the scale invariant behavior $S_3(k_n) \propto 1/k_n$ decreases as
$n$ increases, see
Panels (a) and (b) of Fig~\ref{Fig1}. In contrast, in the case $|c/a|>
1 $ the role of the energy flux becomes irrelevant in the limit of
$n\to \infty$. In this case the properties of the model are completely determined
by the flux of the second integral, see Panels (c) and (d) of
Fig~\ref{Fig1}.  In the sequel we will focus on the region $|c/a| <1$.
The reason for this is that Navier-Stokes turbulence never exhibits
a region in which the energy integral is irrelevant. 

As we discussed, even in the ``physical'' region in which $|c/a| <1$ the sub-leading
contributions (which are effected by the second integral) may influence
the apparent scaling behavior of the leading
scale invariant contributions which are determined by the
energy integral.
In the region $-1<c/a<0$ which is commonly discussed in literature, subleading
contributions lead to period-two oscillations which decrease upon the
increase of $n$. These
introduce additional problems in determining the scaling exponents. A
simple way to eliminate this complication is to consider a
``helicity-free'' forcing chosen such that the flux of the second integral (``helicity")
would vanish identically.
This is easilly achieved by selecting the forcing of the two first
shells. From Eq.~(\ref{eps-delta}) we deduce 
\begin{equation}
  \label{eq:d=0}
  \overline\delta=0\quad {\rm at } \quad c\,p_0+a\,p_1=0,\quad p_2=p_3=\dots=0\ .
\end{equation}
 For a random force  which is Gaussian and $\delta$-correlated in time  
  \begin{equation}\label{ranf}
   \left< f_n(t)f_m(t')\right>=\sigma^2_n \Delta_{nm}\delta(t-t')\,,
  \end{equation}
  one gets 
  \begin{equation}\label{pn}
    p_n=\sigma^2_n\ .
  \end{equation}
For this type of forcing the condition of zero ``helicity'' flux
 (\ref{eq:d=0})
is achieved by choosing the forcing to have the mean-square amplitudes
\begin{equation}
  \label{eq:2d=0}
  c\, \sigma^2_0+a\,\sigma^2_1=0\ .
\end{equation}
Under this condition the period-two oscillations disappear.

The GOY model has some  properties that make it undesirable
for further analytic studies. It is best to exhibit these in comparison
with the new (and we believe superior) model that we refer
to as the ``Sabra" model.

\section{Sabra model: definition and main features}
\label{models}
\subsection{Sabra model}\label{ss-Sabra}
We propose the following equation of motion for the Sabra model:
\begin{eqnarray} \label{sabra}
\frac{d u_n}{dt}&=&i\big( ak_{n+1}  u_{n+2}u_{n+1}^*
 + bk_n u_{n+1}u_{n-1}^*  \\ \nonumber
&& -ck_{n-1} u_{n-1}u_{n-2}\big)  -\nu k_n^2  u_n +f_n\,,
\end{eqnarray}
where for simplicity we assume that the coefficients $a$, $b$, and
$c$ are real. As in the GOY model, conservation of energy in the
inviscid limit is obtained if $a+b+c=0$ .

 The fundamental difference with the GOY model lies in the
 number of complex conjugation operators used in the non linear
 terms. We show in the following that this slight change is
 responsible for a difference in the phase symmetries of the two
 models. As a consequence, the Sabra model will exhibit shorter 
ranged  correlations than the GOY model. Apart from
 this difference, all the calculations described in the 
previous section  remain valid. Both models share the same 
quadratic invariants and one can derive for the Sabra model 
another analog of the 4/5 law. We need to
replace  the definition of the odd orders
correlators (\ref{S2m+1}) according to:
\begin{eqnarray} \nonumber
S_{3}(k_n)&=&{\rm Im}\langle u_{n-1}u_n u^*_{n+1}\rangle \,,  \quad
{\rm (Sabra)} \\
S_{2m+1}(k_n)&=&{\rm Im}\langle u_{n-1}u_n|u_n|^{2(m-1)}u^*_{n+1}
\rangle   \ . \label{Ssabra}
\end{eqnarray}
Note that the shell index $n$ is related to the intermediate shell
involved in the correlation function.

\subsection{Phase symmetry and correlations}
 Let us examine the phase transformation:
\begin{equation}\label{phase}
u_n \to u_n \exp(i \theta_n)\ .
\end{equation}
The equations of motion of both the GOY and the Sabra models remain
invariant under such transformations, provided that the phases
$\theta_n$ are related by:
\begin{equation}\label{invariance}
\begin{array}{ll}
\theta_{n-1}+\theta_{n}+\theta_{n+1}&=0\,,  \qquad {\rm(GOY)}\,, \\
\theta_{n-1}+\theta_{n}-\theta_{n+1}&=0\,,  \qquad {\rm(Sabra)}\ .
\end{array}
\end{equation}
The phases $\theta_n$ can then be obtained iteratively from
 $\theta_1$ and $\theta_2$, namely
\begin{eqnarray}
&&\theta_{1+3p}=\theta_{1},\ \theta_{3p+2}=\theta_{2},  \  
\theta_{3p}=-\theta_{1}-\theta_{2}\,, \   {\rm (GOY)}; \, \nonumber \\
\label{theta1}\\ \nonumber
&&\theta_{n}=\frac{1}{\sqrt{5}}\left[\theta_{1}(\alpha_+^{n-2}
-\alpha_-^{n-2})+\theta_{2}(\alpha_+^{n-1}-\alpha_-^{n-1})
\right]\,,    \\ 
 && \alpha_{\pm}=\frac{1}{2}(1\pm\sqrt{5})\,, \qquad 
\  {\rm (Sabra)}\ .\label{theta2}
\end{eqnarray}
Although Eq.~(\ref{theta2}) has irrational numbers, it is
easy to check that 
\begin{equation}
\theta_n=r_n\theta_1+ s_n \theta_2 \,,
\end{equation}
where $r_n$ and $s_n$ are integer numbers which grow exponentially
with $n$.

Note that phases $\theta_1$ and $\theta_2$ satisfy the
equations of motion
\begin{equation}\label{random}
\frac{d \theta_1}{d t}=0\,, \quad \frac{d \theta_2}{d t}=0\,,
\end{equation}
and they can be randomized by any small external forcing.  It means
that any correlation functions which contain the phases $\theta_1$,
$\theta_2$ or both phases must be zero. In our direct numerical
simulations, see below, we confirmed that this is indeed the case. In
the Sabra model there is only one nonzero 2nd order structure
function.  All nondiagonal correlation functions vanish in the Sabra
model
\begin{equation}\label{nond}
S_2(k_n,k_m)=\left< u_n u^*_m \right>=0 \quad  n\ne m~ {\rm (Sabra).}
\end{equation}
This is not the case for the GOY model for which there are
correlations between shells separated by multiples of three:
\begin{equation}\label{nond0}
S_2(k_n,k_{n+3p}) \ne 0 \quad {\rm (GOY)}\ .
\end{equation}
The relative simplicity of the Sabra model is seen also with regards
to higher order structure functions. The Sabra model has only one
non-zero 3rd order structure function $S_3(k_n)$ that couples three
consecutive shells as defined by Eq.~(\ref{Ssabra}). All other 3rd
order structure functions vanish by averaging over the random phases
$\theta_1$ and $\theta_2$.  In contrast, in the GOY model there exists
an infinite double set of nonvanishing correlation functions of 3rd
order with given $n$. These are
\begin{equation} \label{nond1}
\left< u_n u_{n+3p}u_{n+1+3q}\right>\ne 0 \,, 
\quad{\rm (GOY)}\ .
\end{equation}
The same phenomenon occurs also for higher order correlation
functions. In the Sabra model the number of non-zero correlation
functions with finite $n$ is much smaller then the corresponding
functions in the GOY model, making it more convenient for theoretical
analysis.

To conclude this section we formulate a ``conservation law'' that
determines which correlation functions of the Sabra model are
non-zero. Introduce a quasi-momentum $\kappa_n$ for $n$--shell by
\begin{equation}\label{nond2}
\kappa_n\equiv \alpha^n\,,
\end{equation}
where $\alpha$ is the golden mean, $\alpha^2=\alpha+1$. One can check
that in the Sabra model the only non-zero correlation functions
satisfy the following conservation law: {\em the sum of incoming
  quasi-momenta (associated with $u$) is equal to the sum of outgoing
  quasi-momenta (associated with $u^*$)}.
\subsection{Additional properties}
In this subsection we show that the
Sabra model exhibits the properties of the GOY model which were 
revealed in Subsect.~\ref{ssect:2.2}.

With this aim we compute from Eq.~(\ref{sabra}) the time derivative of
$S_2(k_n,t)$:
\begin{eqnarray}
&& {dS_2(k_n)\over  dt}= 2 {\rm Re } \Big \langle 
\frac{d u_n(t)}{dt} u_n^*(t) \Big \rangle \\
&=&-2 {\rm Im }\Big[ak_n \langle u_n^*u_{n+1}^*u_{n+2}
\rangle+bk_n\langle u_{n-1}^*u_n^*u_{n+1}\rangle
 \nonumber \\
&& -ck_{n-1}\langle u_{n-2}u_{n-1}u_n^* \rangle\Big]-2\nu k_n^{2}
\langle u_n u_n^* \rangle+p_n , \nonumber
\end{eqnarray}
where the forcing contribution $p_n$ was defined in Eq.~(\ref{pump}).

With the definition (\ref{Ssabra}) of $ S_3(k_n)$ this translates to
the balance equation (\ref{flux}) derived for the GOY model. Note that
these two models differ in the definitions of $ S_3(k_n)$:
Eq.~(\ref{S2m+1}) for the GOY model and Eq.~(\ref{Ssabra}) for the
Sabra model.  Clearly, $ S_3(k_n)$ in the Sabra model has the
same form~(\ref{al}) as in the GOY model and all the features of the GOY
model discuused in Subsect.~\ref{ssect:2.2} are relevant for the Sabra
model as well.  In particular one may eliminate the period-two
oscillations by a proper choice (\ref{eq:d=0})  or (\ref{eq:2d=0})  of
the forcing.

The reader should note however that in the case of the GOY model the
second and the third order structure functions have additional long
range correlations which do not appear in the balance equation.  This
is a sickness of the GOY model that is eliminated in the context of
the Sabra model, where what you see is what exists.  Note also
  that the long range correlations for the GOY model exists between
  shells separated by multiples of three (see, for example
  Eqs.~(\ref{nond0}), (\ref{nond1})). These correlations are
  responsible for period-three oscillations in scaling plots of the GOY model.  
These annoying oscillations are absent in the Sabra model by 
construction. Thus after elimination of the period-two oscillations
(using ``helicity-free'' forcing) one finds scale invariant behavior of
the structure functions almost from the very beginning of the inertial 
interval. 
\section{Aspects of the numerical integration: stiffness, forcing and dissipation}
The numerical investigation of the Sabra model, as of any other stiff
set of differential equations, calls for some care. We therefore
dedicate this section to a discussion of the issues involved. A reader
who wishes to consider the results only can skip this section and read
the next one.
\subsection{Stiffness}
The main difficulty in integrating a shell model stems obviously from
the stiffness of the system {\it i.e.} we are concerned with a wide
range of time scales in the system. Within the inertial range, the
equation is dominated by the non linear terms so that the natural time
scale (in the Kolmogorov approximation) of the $n$th shell scales as:
\begin{equation}
\tau_n \sim \frac{1}{k_n u_n} \propto {1 \over k_n^{2/3}}\ .
\end{equation}
Within the viscous range however, the dominant term is the viscous one
and if the $n$th shell lies in this subrange, its natural time
becomes:
\begin{equation}
\tau_n \sim \frac{1}{\nu k_n^2} \ .
\end{equation}
We can now estimate the global stiffness of the system by quoting the
ratio of the extremal time-scales: 
\begin{eqnarray} \frac{\tau_1}{\tau_N} &\sim&
\frac{\tau_1}{\tau_{n_d}} \frac{\tau_{n_d}}{\tau_N}\sim \left(
  \frac{k_d}{k_1}\right)^{2/3} \left( \frac{k_N}{k_d}\right)^2 \\
\nonumber &&\propto 
\lambda^{2[N+2(N-n_d)-1]/3} \ .  
\end{eqnarray} 
The global stiffness of the system thus depends both on the total
number of shells $N$ and on the width of the viscous region.  Most of
the results published in the literature are obtained with 22 shells, a
forcing restricted to the first shell and a viscous boundary beginning
about the 18th shell. In this typical case, we have $\tau_1/\tau_N
\sim 6.6 \times  10^5$. In this paper we typically use $N=34$ with about 6
 shells in the viscous range. For this choice $\tau_1/\tau_N \sim
10^9$.

To deal with this stiffness we chose from the library SLATEC
\cite{SLATEC} the backward differentiation routine DDEBDF
\cite{DDEBDF}. This routine is specially dedicated to very stiff problems.
Although rather fast, its precision is not exceptional and it is
rather sensitive to functions which are not sufficiently smooth. In
cases of failure of the backward differentiation routine, the code
switches automatically to a $4/5$th order Runge-Kutta
algorithm. Both routines adapt their step-size to fulfill a prescribed
precision requirement. The backward differentiation routine adapts in
addition its order between 1 to 5.
\subsection{Random forcing} 
We generate the random forcing to guarantee zero mean value of the
velocity. We use a time correlated noise, with a correlation time
chosen to be the natural time scale at the forcing shell:
$\tau=1/(k_{n_L}u_{n_L })$. Denoting the forcing term $f$, in case
of an exponential correlation, the evolution of $f$ is ruled by the
equation
\begin{equation}
\frac{d}{dt} f =-\frac{f}{\tau} +\eta \,,
\end{equation}
where $\eta$ is an uncorrelated noise. The presence of this new
equation in the system could in principle make the integration more
cumbersome. Fortunately, the system being stiff, the typical time step
used in the integration is very small compared with the forcing time
scale $\tau$ (six orders of magnitude in a typical calculation with 22
shells). This allows us to integrate f separately with a first order
scheme. In the code, the forcing is updated at each new call of the
integrator. The Gaussian exponentially correlated random forcing is
computed (after proper initialization) according to a first order
scheme proposed by Fox et {\it al}\cite{Fox}:
\begin{equation} f(t+dt)= f(t) E + \sigma
\sqrt{-2(1- E^2) \log(\alpha)} \exp\left(i 2\pi \beta \right) 
\end{equation} where $E=\exp
( -dt/\tau)$, $\sigma$ is the standard deviation of $f$ and $\alpha$ and
$\beta$ two random numbers between 0 and 1.

\noindent
\subsection{Dimensional Analysis}
\begin{figure}
\epsfxsize=8truecm  
\epsfbox{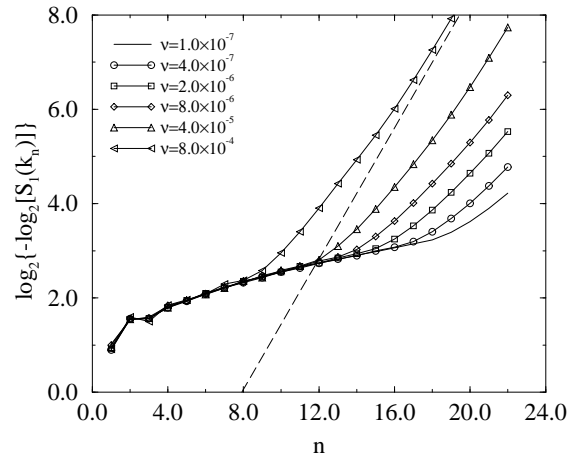}
\caption
{Modulus of $u_n$ in the Sabra model obtained by integration over 500
turn-over time scales with values of the viscosities as shown in the
figure.  The dashed line represents the expected asymptotic behavior
in the deep viscous regime. The slope of this line is given by
equation (39).}
\label{Figtwo}
\end{figure}

For the purpose of our numerical fits we consider, following \cite{Schorg95},
the dissipative boundary $n_d$, where the
dissipative term balances the non linear term. At this boundary 
$k_d u_{n_d}^2 $ is of the order of $\nu k_d^2 u_{n_d}$.
In the viscous range $n>n_d$ one can guess a generalized exponential form:
\begin{equation}
u_n \sim k_n  \exp \left[- \left(
\frac{k_n}{k_{d}}\right)^x\right]\ , \label{expdam}
\end{equation}
where  \cite{Schorg95}
\begin{equation}\label{def-x}
x=\log_\lambda \frac{1+\sqrt{5}}{2}\ .
\end{equation}
We have studied the influence of the width of the viscous range on
this exponential behavior.  The results obtained for a system of 22
shells with various viscosities are summarized in Fig.~\ref{Figtwo},
where we can see that the scaling behavior in the viscous range
approaches slowly the asymptotic prediction. In the case of the
largest viscosity used $\nu=8\times 10^{-4}$, we note that the asymptotic
behavior starts at $n \simeq 15$ while $n_d \simeq 9$. We can then
consider that this width of 6 shells is the minimal one needed to
properly describe the viscous range.

In the inertial interval dimensional reasoning leads to K41 scaling:
$u_n \sim (\bar\epsilon / k_n)^{1/3}$.   This  formula  may be
matched with (\ref{expdam}):
\begin{equation}
u_n \sim u_{n_L} \left( \frac{k_{n_L}}  {k_n}\right)^{\frac{1}{3}} 
\left[1+ \left(\frac{k_n}{k_d} \right)^{\frac{4}{3}} \right]
\exp\left[- \left(\frac{k_n}{k_{d}}\right)^x \right] \ , \label{dimsol}
\end{equation}
where $ u_{n_L} \sim \sqrt{f/k_{n_L}}$ and $ k_d \sim \ ( f^3 / \nu^6
k_{n_L})^{1/8}$.  We will see that although the actual values of the
exponents change due to multi-scaling, the form of the solution is
rather close to reality, and Eq.~(\ref{dimsol}) is a good starting
point for numerical fits.
\section{Numerical Simulations: Results}
A careful determination of the scaling exponents is a delicate issue.
With an infinite inertial range, we expect pure scaling laws. Despite
its large size, the inertial range that we have in shell models
remains finite. The most widely used method to determine the exponent
is based on a linear regression or on the determination of a local
slope\cite{SL95PRL} in log-log scale. In these methods one needs a
criterion to choose the fitting range. The uncertainty in the scaling
exponents comes obviously from the quality of the regression but also
largely from the number of shells taken into consideration. We want to
make the point here that these methods are not reliable, giving rise
to a lot of confusion in the literature. One needs to fit a whole
function to the inertial and dissipative ranges simultaneously to
achieve reliable estimates of the exponents in the inertial range.

The definition of the scaling exponents can be matter of choice of the
statistical object.  Our preferred definition is (\ref{S2m}) and
(\ref{Ssabra}) for even and odd exponents respectively. Two
alternative choices were widely used in the literature, respectively
based on the modulus of the velocity and on the energy flux:
\begin{eqnarray}\label{tildeSp}
&&\tilde S_q(k_n)=\big \langle |u_n|^{q}   \big \rangle \,, \\ \label{hatSp}
&& \hat S_q(k_n)= \big \langle |\Sigma_n|^{q/3}\big \rangle \\
&=&\big \langle \left| {\rm Im} \left[a \lambda \,u_{n}u_{n+1}u_{n+2}^* 
- c \, u_{n-1}u_n u_{n+1}^* \right] \right|
^{q/3} \big \rangle \ .\nonumber
\end{eqnarray}
The latter definition allowed for a higher numerical precision in the
context of the GOY model because the energy flux is not affected
either by the genuine dynamical oscillation (due to to the helicity
flux) or by the period three oscillations. Beyond these different
definitions of the statistical objects, we can also modify the
definition of the scaling exponents themselves. In the framework of so
called ``Extended Self Similarity", instead of writing $S_q(k_n)=A
k_n^{-\zeta_q}$ one assumes a scaling relation between the structure
functions of order $q$ and of order 3: $S_q(k_n)=A[S_3(k_n)]^{{\tilde
    \zeta}_q}$.

These different definitions give a priori different sets of scaling
exponents. An efficient comparison is however difficult to set up in
the case of the GOY model because of the various oscillations
polluting the data. Moreover none of the techniques described so far
took explicitly into account the finite size effects. The fitting
procedure that we describe now is a first attempt to do so, and one of
the results is that the exponents are universal, independent (for
given parameters) of the choice of the statistical object.

In light of the interpolation formula (\ref{dimsol}), and encouraged by
the fact that the dissipative, stretched exponential behavior is
rather nicely obeyed, we fit all our spectra to the following fit
formula:
\begin{equation} F_q(k_n)={A_q\over  k_n^{\zeta_q}}
 \left(1+\alpha_q \frac{k_n}{k_{d,q}}
\right)^{\mu_q} \exp\left[-\left(\frac{k_{n}}{k_{d,q}}\right)^x\right]
\ . \label{fit} \end{equation} 
This guarantees the right behavior at both asymptotics. Note that we
don't make any hypothesis on the form of the transition between the
power law and the dissipative regimes. In fitting we minimize the
following error function:
\begin{equation}
{\cal E}= \sqrt{\sum_n \Big[1- \frac{\log F_q(k_n)}{\log
S_q(k_n)}\Big]^2 } \ . \label{minimize}
\end{equation}
Here $S_q$ refers to the numerically obtained structure function. We
use the same fit formula for all the three definitions of statistical
objects.  The sum in (\ref{minimize}) was computed over the whole
range except the two first shells and the two last shells in order to
limit the effect of the boundaries. It turns out that the minimum
found in this procedure is sharp (as a function of $\zeta_q$) {\em
  provided} that we have a good fit of the $S_q$ over its whole range.
To estimate the relative error in the scaling exponents $\zeta_q$ we
arbitrarily computed the values of $\zeta_q$ that agree with values of
${\cal E}$ that are twice the minimum value. These are the errors
reported in all the tables below.

\begin{figure}
\epsfxsize=8.5truecm
\epsfbox{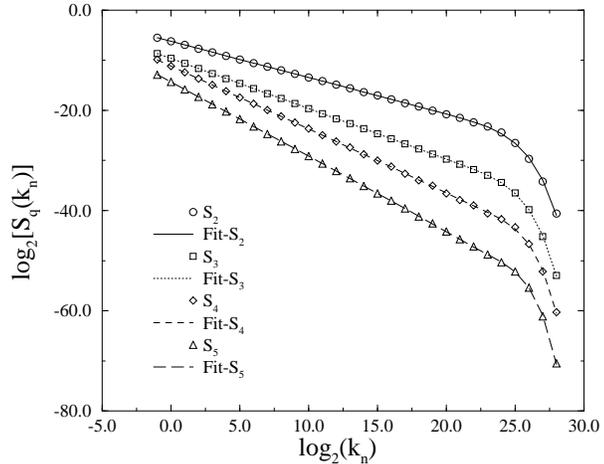}
\caption
{Log-log plot of the structure functions $S_2(k_n)$ to $S_5(k_n)$
vs. $k_n$ and of the results obtained using the fitting formula
(44). The structure functions are represented by the symbols
and the fits by the lines.}
\label{fitS2S5}
\end{figure}

In all our simulations we used the parameter values $a=1$, $b=c=-0.5$
and $\sigma_1/\sigma_0=0.7$. This choice 
eliminates the flux of helicity and correspondingly the period-two
oscillations in the scaling plots. Typical fits for the structure functions from
$S_2$ to $S_5$ for simulations with 34 shells ($\nu=4\times 10^{-11}$,
$\sigma_0=5\times10^{-3}$,) are shown in Fig. \ref{fitS2S5}.
In Table~\ref{tab1} we present the computed scaling exponents
associated with three different definitions of $q$-order
  correlation functions.  These results offer a very
  strong indication that the three scaling exponents of $q$-order
  correlation functions (with given $q$) are all the same.

\begin{figure}\label{ess}
\epsfxsize=8.3truecm
\epsfbox{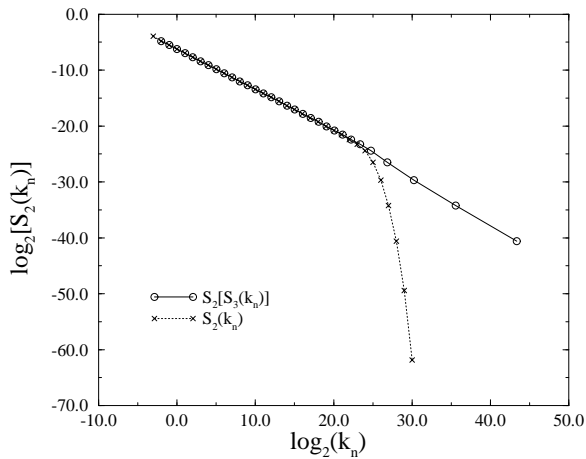}
\caption
{Log-log plot of $S_2(k_n)$ vs. $k_n$ and vs. $S_3(k_n)$ respectively.
  This plot shows that at least for this model, and when the accuracy
  is sufficiently high, ESS is quite useless in increasing the
  effective range of power law behavior.}
\label{Fig4}
\end{figure}

On the other hand, we can make the point that ``Extended
Self-Similarity" (ESS)\cite{93Ben} in its standard usage does not seem
be a useful approach in the present context for computing more
accurate scaling exponents. In Fig.~\ref{Fig4} we present $S_2(k_n)$
both as a function of $k_n$ and as a function of $S_3(k_n)$. Even
though superficially the ESS way of plotting seems to yield a longer
linear plot, a careful examination shows a break in the inertial range
scaling which occurs precisely at the crossover to dissipative
behavior. We gain nothing from ESS for this model.

Nevertheless, for the limited aim of computing a precise value of
$\zeta_2$, we can make use of the ESS idea provided that we fit the
whole range. To do this we have to impose additional information on
the fitting function. For the case of $\zeta_2$ we can employ the
information contained in the balance equation (\ref{flux}), closing it
with the ansatz
\begin{equation}\label{a12}
 S_2(k_n)=A_2 |S_3(k_n)|^{\zeta_2}\ .
\end{equation}
Using (\ref{a12}) and introducing $z_n=-k_n S_3(k_n)$, we can rewrite
(\ref{flux}) as:
\begin{equation}\label{a13}
z_{n+1}=z_{n-1}+b(z_{n-1}-z_n)-\Big(k_n/k_*
\Big)^{2-\zeta_2}|z_n|^{\zeta_2},
\end{equation}
where $k_*=(\nu A_2)^{1/(\zeta_2-2)}$ and $a=1$.

Given $z_0$ and $z_1$, one can iteratively calculate $z_n$ and,
consequently, $S_3(k_n)$ and $S_2(k_n)$ in the range of $k_n$, for
which the ESS ansatz is valid with reasonable accuracy.  Assuming for
simplicity $z_0=z_1$, the values of $z_n$ are defined by 3 free
parameters: $z_0, A_2, \zeta_2$.

\begin{table}
\begin{center}
\begin{tabular}{||c|c|c|c||}
q &$S_q$ &$\langle|u_n|^q\rangle$ &$\langle|
\Sigma_n|^{q/3}\rangle$ \\
 \hline  
1 &\  &$0.393\pm 0.006$  &$0.393\pm 0.007$ \\
 \hline  
2 &$0.720\pm 0.008$ &$0.720\pm 0.008$  &$0.719\pm 0.007$ \\ 
\hline  
3 &$1.000\pm 0.005$ &$1.003\pm 0.009$  &$1.000\pm 0.005$ \\ 
\hline  
4 &$1.256\pm 0.012$ &$1.256\pm 0.012$  &$1.249\pm 0.003$\\ 
\hline  
5 &$1.479\pm 0.006$ &$1.488\pm 0.013$  &$1.477\pm 0.004$\\ 
  \hline 
6 &$1.706\pm 0.015$ &$1.706\pm 0.015$  &$1.691\pm 0.006$\\ 
\hline 
7 &$1.901\pm 0.010$ &$1.910\pm 0.020$  &$1.893\pm 0.010$\\ 
 \end{tabular}
\end{center}
\caption {Summary of the scaling exponents computed with a model of 
34 shells }\label{tab1}
\end{table}
\noindent

As an example, we applied this procedure to the numerical data
calculated with $a=1$, $b=-0.5$, and $\nu =4 \times 10^{-11}$. The
values of fitting parameters corresponding to the global minimum of
the functional $\cal E$ (\ref{minimize}) are $z_0=0.00126$,
$A_2=1.80$, and $\zeta_2=0.728$. To estimate the accuracy of the
chosen fit parameters we have studied the dependence of $\cal E$ on
the deviation $\delta \zeta_2$, $\delta A_2$, and $\delta z_0$ from
their optimal values with two other parameters fixed at the optimal
values. As before we define the error bar for each parameter interval
for which $\cal E$ takes on values which are twice the value at the
minimum.  With this definition $z_0=0.00126\pm 0.00002$, $A_2=1.80\pm
0.06$, and $\zeta_2=0.728\pm 0.002$.

The accuracy reached here is higher than in the procedures described
above. Most of the errors in the fit appear from the crossover region
from power law to exponential decay.  The analytically calculated
$S_2(k_n)$ and $S_3(k_n)$ near the onset of the viscous range are very
sensitive to the value of $\zeta_2$. Therefore employing an adequate
fit in this region (which uses additional {\em a priori} information
contained in the balance equation) allows one to be more accurate.
Note that we do not have such simple balance equations for higher
order correlation functions and therefore a generalization of the
procedure for higher orders is not available.

\section{Tests  of the statistical quality of the numerical data} 
\label{6}
In evaluating the scaling exponents $\zeta_q$ one has to make
sure that the structure functions $S_q(k_n)$ are calculated properly.
This means that (i) the averaging time is sufficient for the
representative statistics, and (ii) the numerical procedure produces an
accurate realization $u_n(t)$.
\subsection{The PDF test for the averaging time}
In intermittent statistics one may need to wait rather long times
before the appearance of rare events which nevertheless
  contribute significantly to the statistics of $q$-order structure
  functions of $n$-shells. This issue was carefully discussed by
  Leveque and She \cite{97LS} in numerical simulations of the GOY model.
They considered the waiting time $T{n,q}$ which are needed to evaluate
safely $q$ order correlations of $n$ shells. They argued that
times of the order of $5\times 10^9$ turnover times
of the $n$-shell are required for $q\approx 15$.

In the beginning of this subsection we estimate analytically the
waiting time $T_{n,q}$  which is needed to observe, say, 100
  events contributing to $S_{2q}(k_n)$.  This is done using the 
  probability $W_{n,q}$ to observe one rare event in which the value of
the velocity $u_n$ hits the range that contributes mostly to the
statistics of $S_{2q}(k_n)$. Denoting by $\tau_n$ the decorrelation time
on the $n$th shell we estimate
\begin{equation}\label{b22} 
T_{n,q} \sim 100\tau_n /  W_{n,q}\ .
\end{equation}
The probability $W_{n,q}$ may be related to the PDF of the velocity at
the $n$th shell, $P_n(u)$. For the sake of this estimate we take
$P_n(u)$ as a stretched exponential.   We do not imply that this
distribution function is realized in this model (in fact we know that
it is not consistent with multiscaling). We use it only for the sake
of an order of magnitude analytical estimate of the waiting time.
Consider 
\begin{equation}\label{a2}
P_n(v)=C \exp[-|v|^\delta] \,,
\end{equation}
where $v$ is dimensionless velocity $v=u/u_0$, $u_0$ is a
characteristic velocity, $u_0^2\simeq S_2(k_n)$ and 
$C$ is a normalization constant.  One computes $S_{2q}(k_n)$ as
\begin{equation}\label{a1}
S_{2q}(k_n)=  u_0^{2q } \int_{-\infty}^{\infty} v^{2q} P_n(v)dv \ .
\end{equation}
The integrand in  (\ref{a1}) has a maximum at 
$v=v_q$, where
\begin{equation}\label{a4}
 v_q= (2q/\delta)^{1/\delta} \ .
\end{equation}
From (\ref{a2}) we can estimate the probability that $v$ will
  attain a value within an interval of order of $\sqrt{q}\sim 1$
  around $v_q$, which was denoted as $W_{n,q}$. This interval of $v$
  values contributes maximally to $S_{2q}$. Namely,
\begin{equation}\label{a5}
W_{n,q}\sim P_{n}(v_q)=   C \exp[-2q /\delta] \,                \ .
\end{equation}
Equation (\ref{a5}) leads to the estimate
\begin{equation}\label{a6} 
T_{n,q} \sim 100  \tau_n \exp(2q/\delta)\,,
\end{equation}
where $\tau_n$ is a characteristic decorrelation time for $n's$ shell.
The time $T_{n,q}$ is exponentially large.  For instance, for $\delta=1$
and $2q=10$, the averaging time required for accurate measurement of
$S_{10}(k_n)$ is of the order of
\begin{equation}\label{a7} 
T_{n,q}\simeq 100 e^{10} \tau_n \simeq 2\times10^6 \tau_n \ .
\end{equation}

\begin{figure}
\bbox{\epsfxsize=4.3truecm\epsfbox{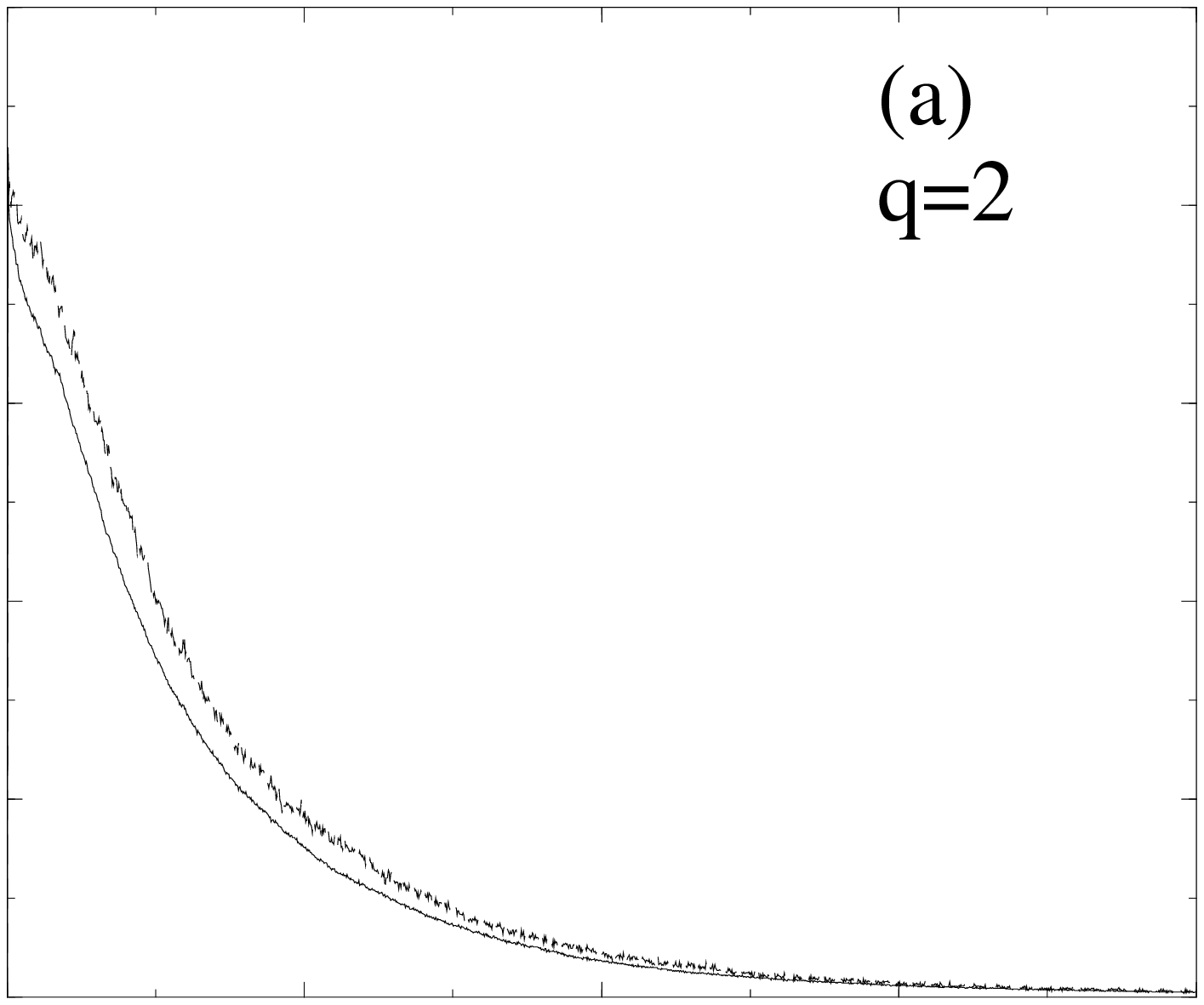}
\epsfxsize=4.3truecm\epsfbox{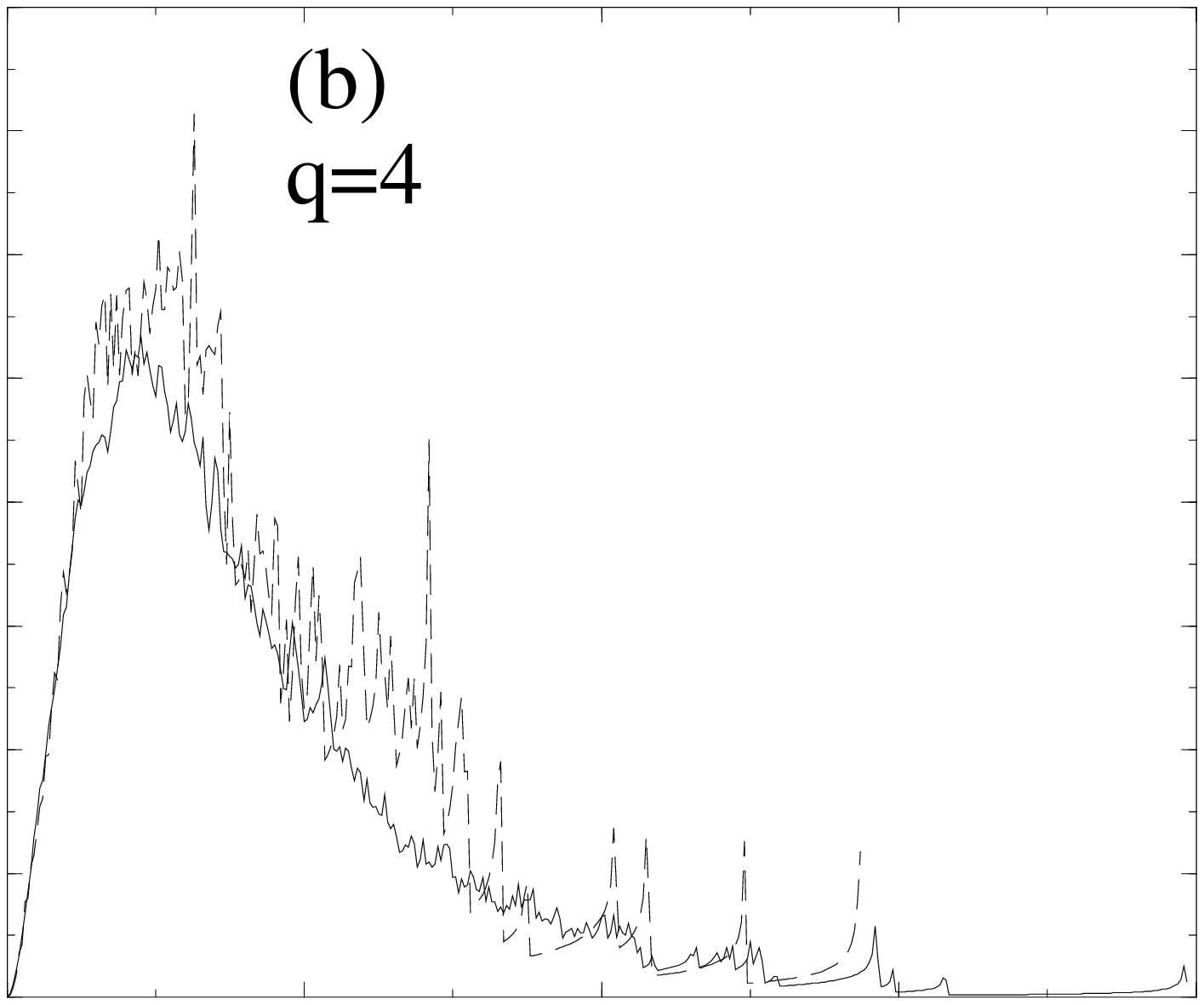}}

\bbox{\epsfxsize=4.3truecm\epsfbox{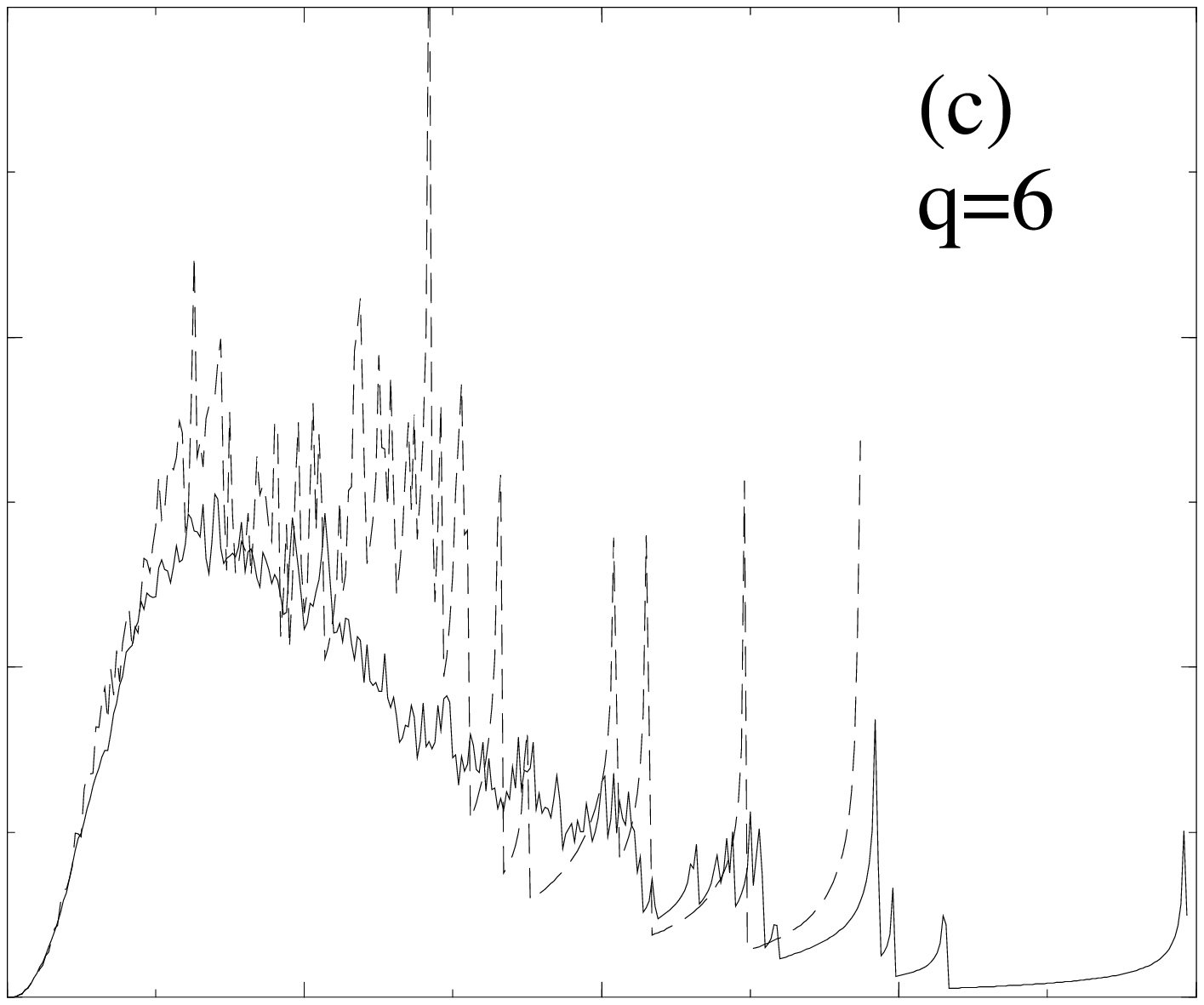}
\epsfxsize=4.3truecm\epsfbox{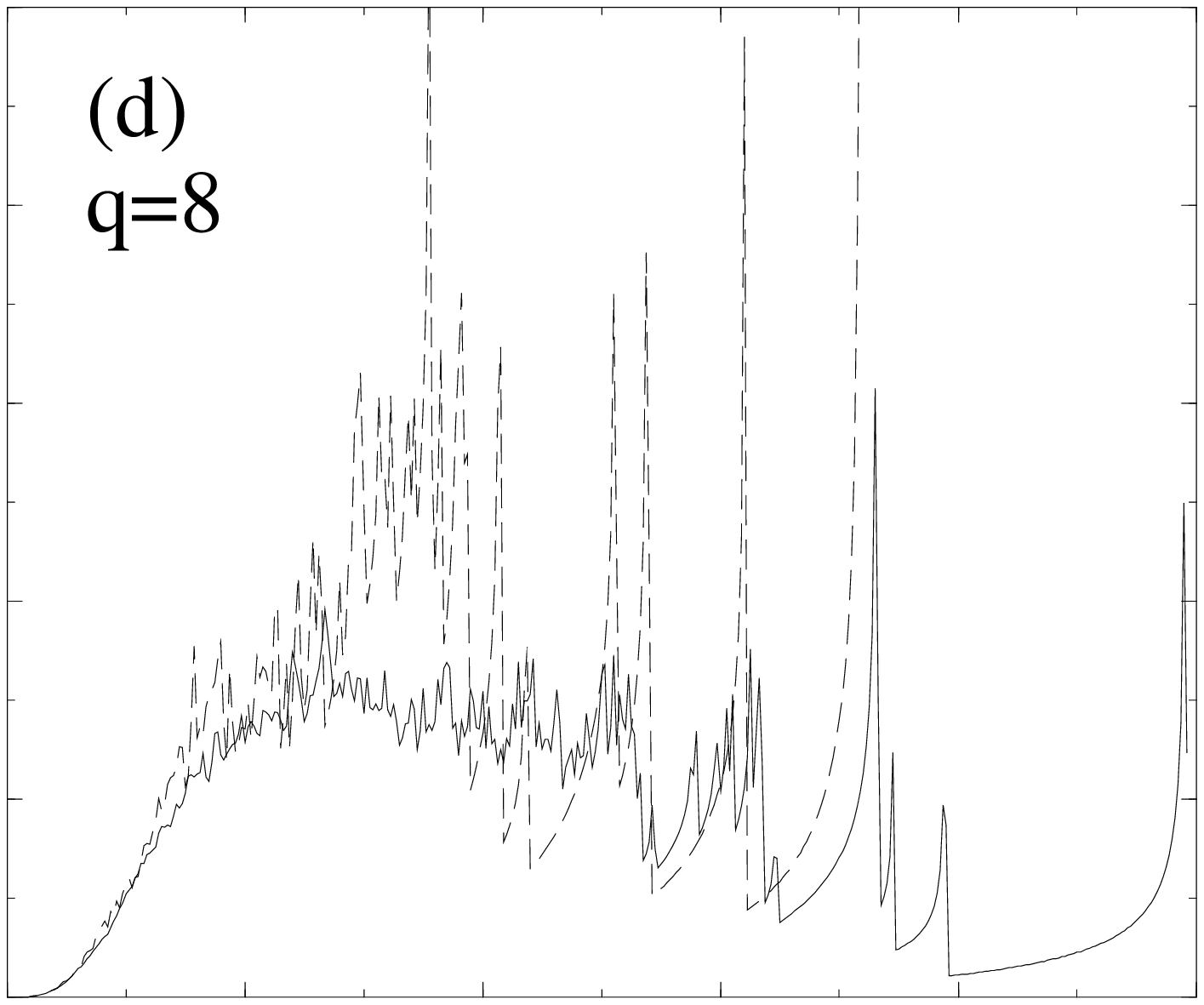}}

\bbox{\epsfxsize=4.3truecm\epsfbox{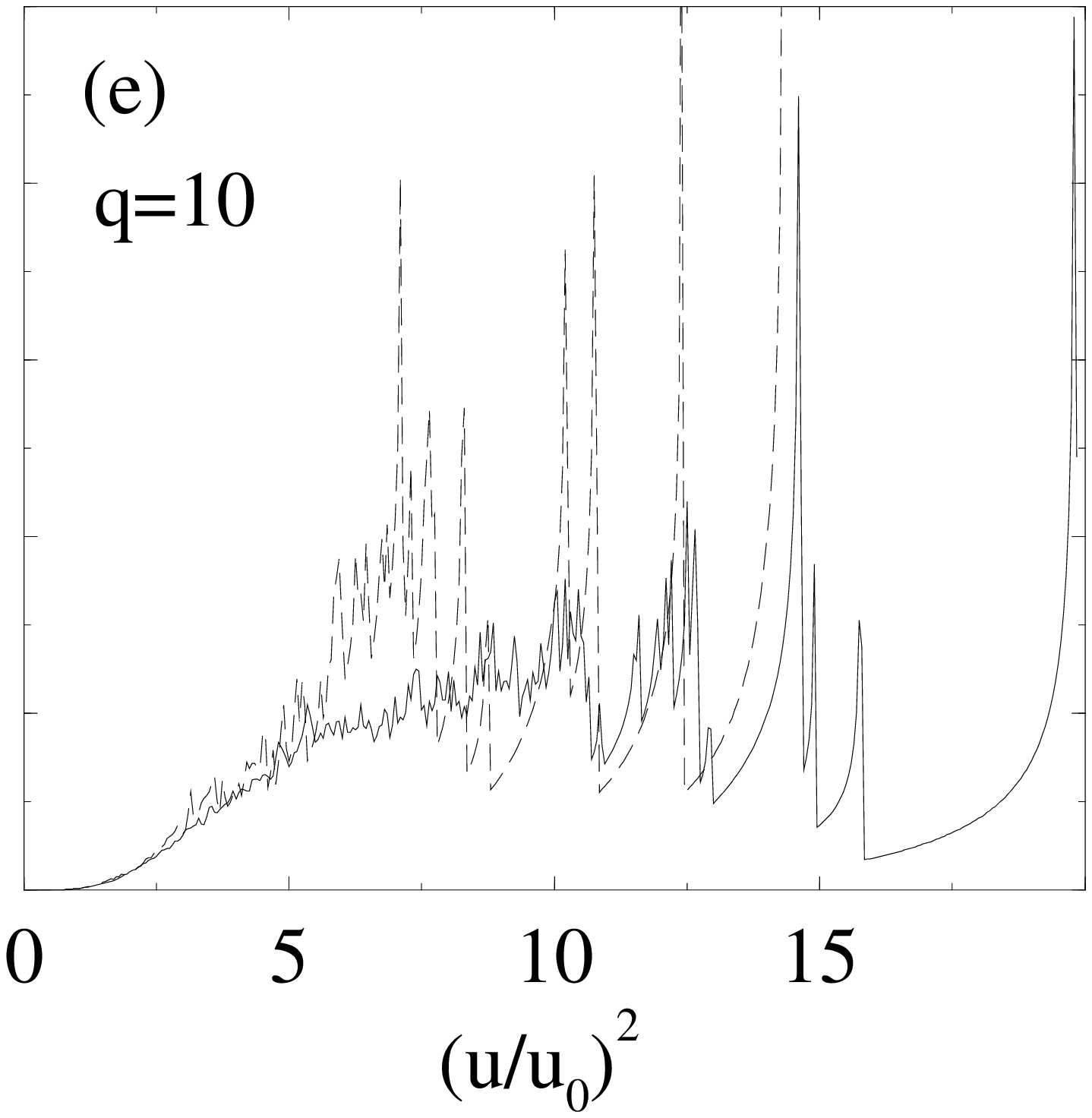}
\epsfxsize=4.3truecm\epsfbox{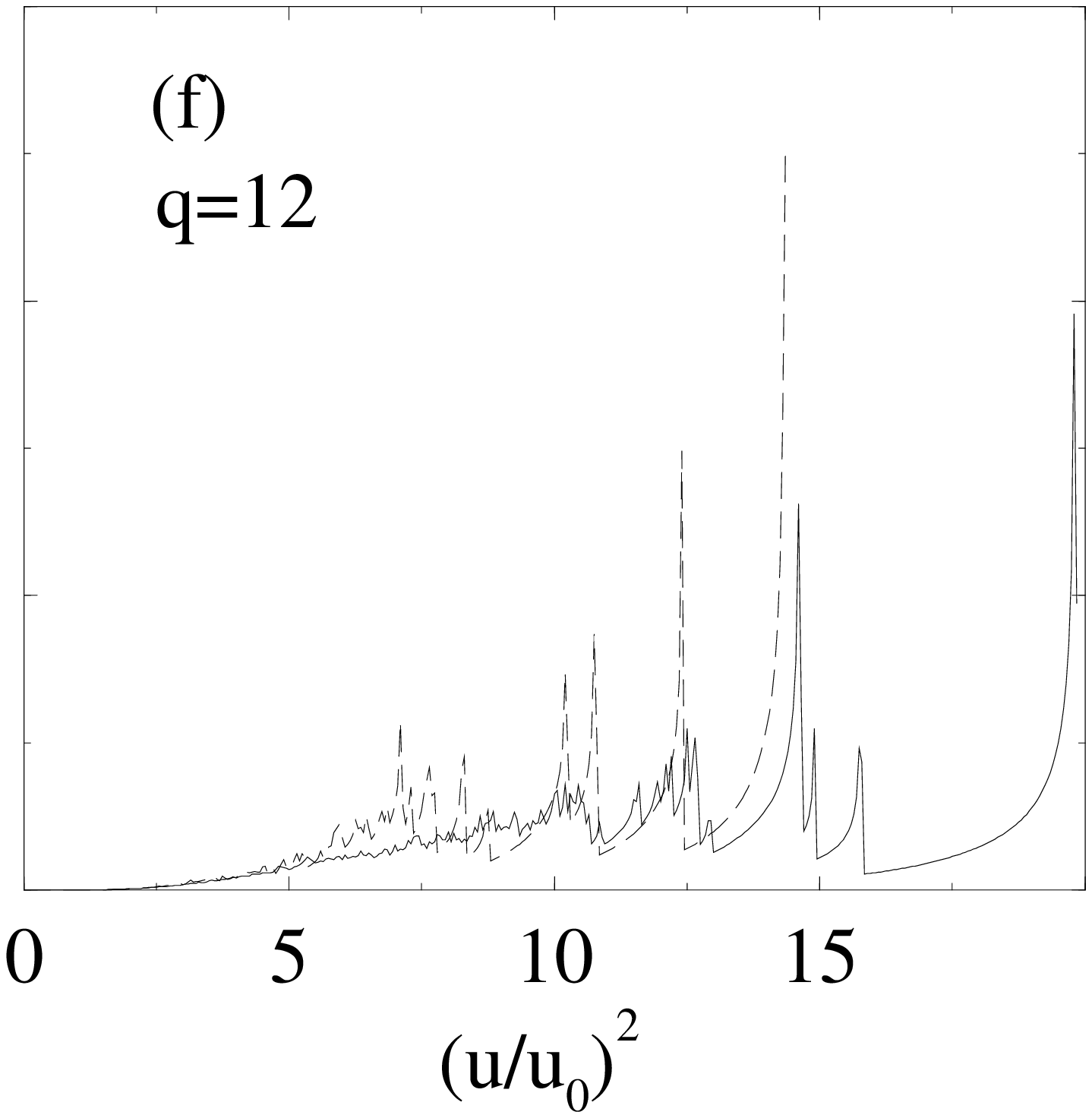}}
\caption{Plots of $(u/u_0)^q  P_3 ((u/u_0)^2)$ for the
third shell with different values of $q$ as shown in the figures.  In every
figure results are presented for 6250 (solid line) and 625 (dashed line)
turnover times $\tau_3$. Already $S_8$ is not accurate even with the 
 longer run.}
\label{XX}
\end{figure}
\begin{figure}
\bbox{\epsfxsize=4.3truecm\epsfbox{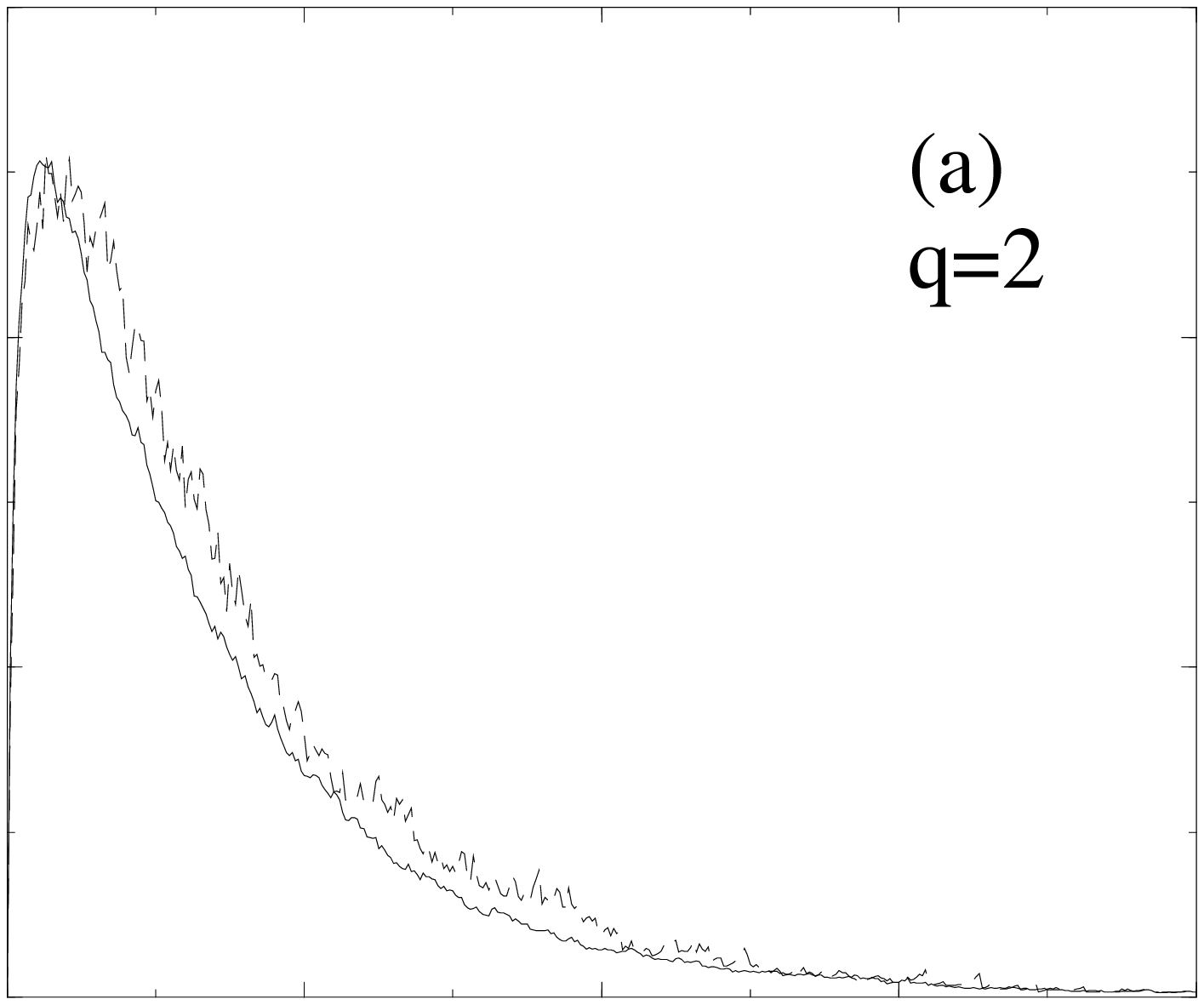}
\epsfxsize=4.3truecm\epsfbox{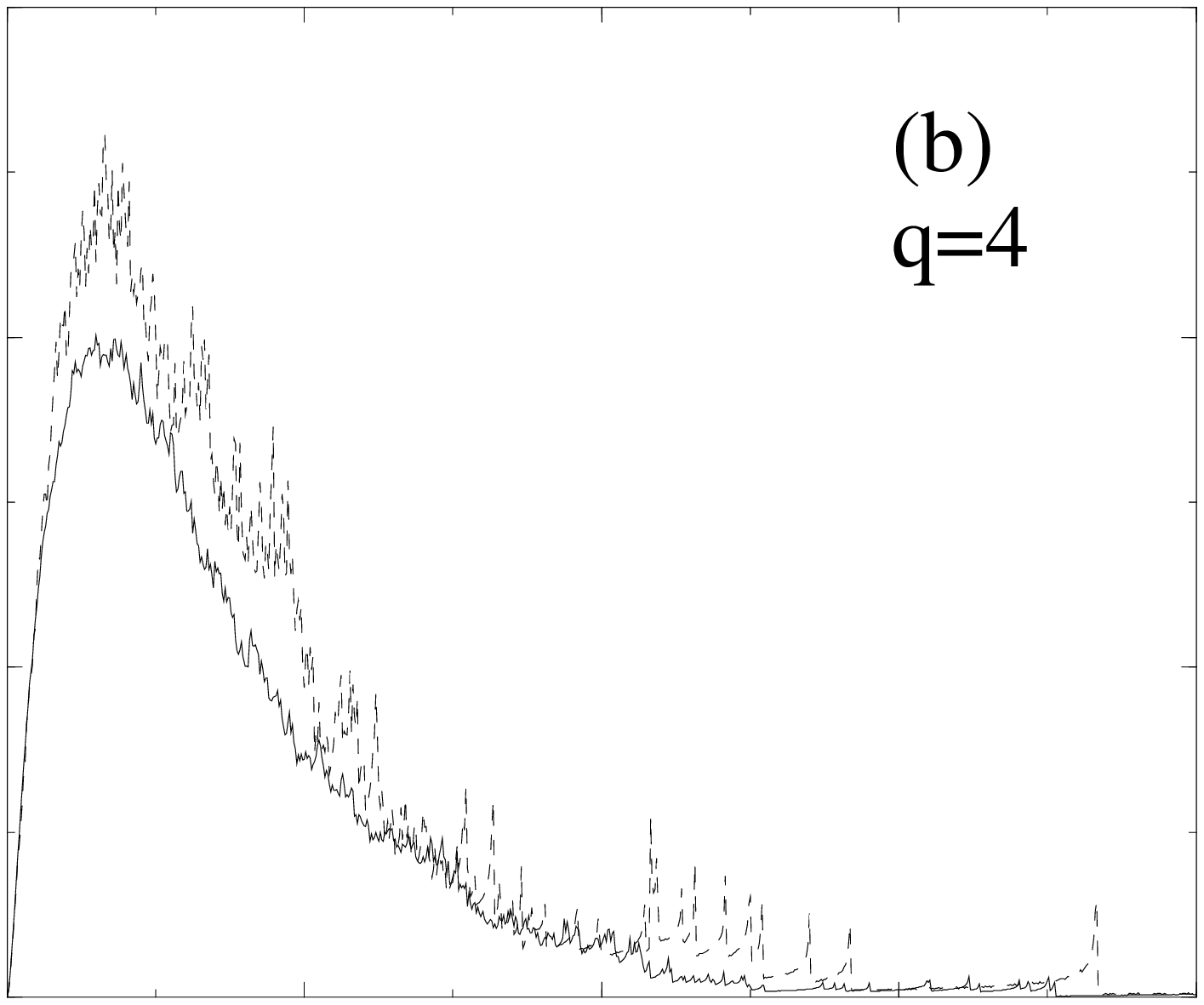}}

\bbox{\epsfxsize=4.3truecm\epsfbox{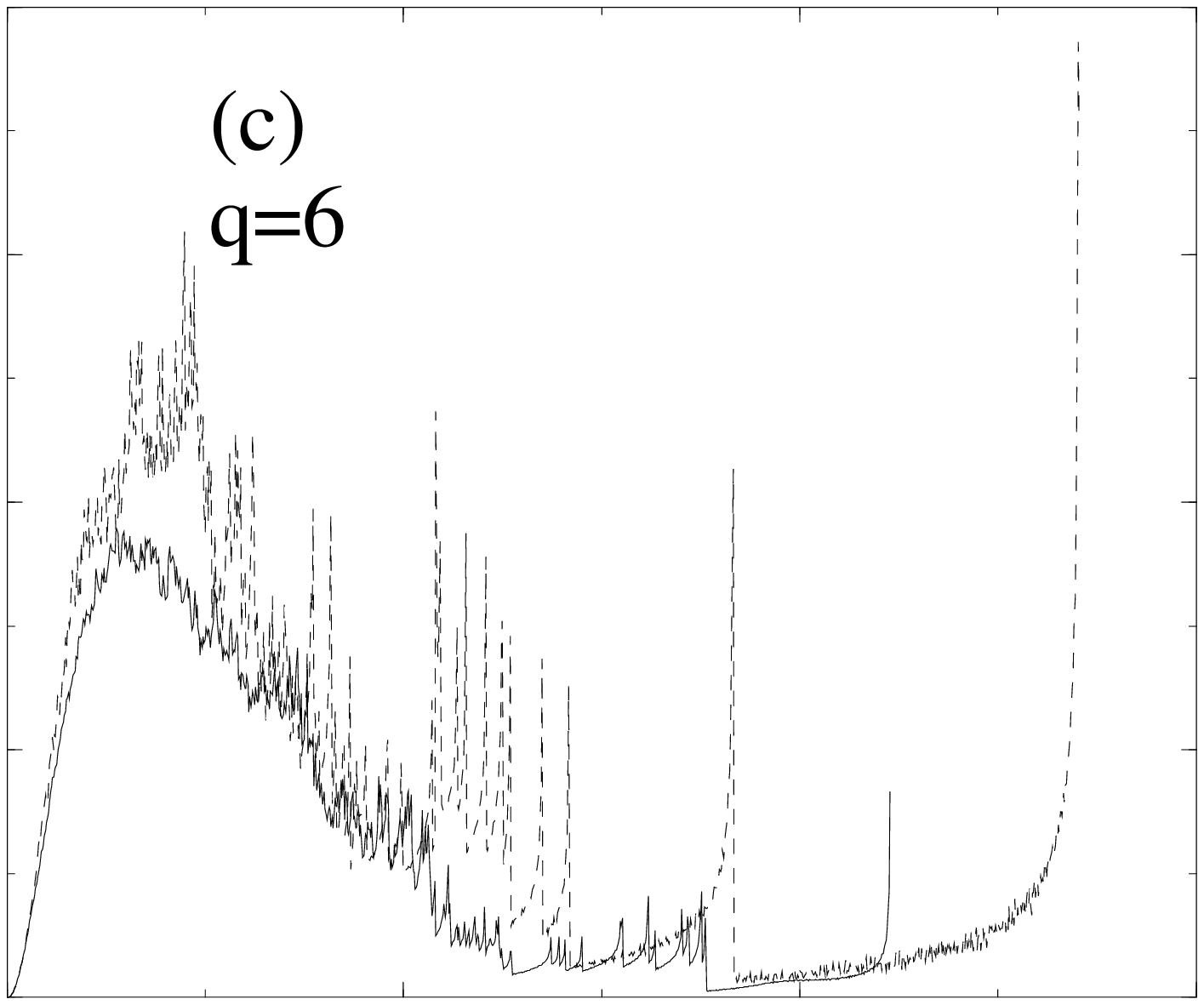}
\epsfxsize=4.3truecm\epsfbox{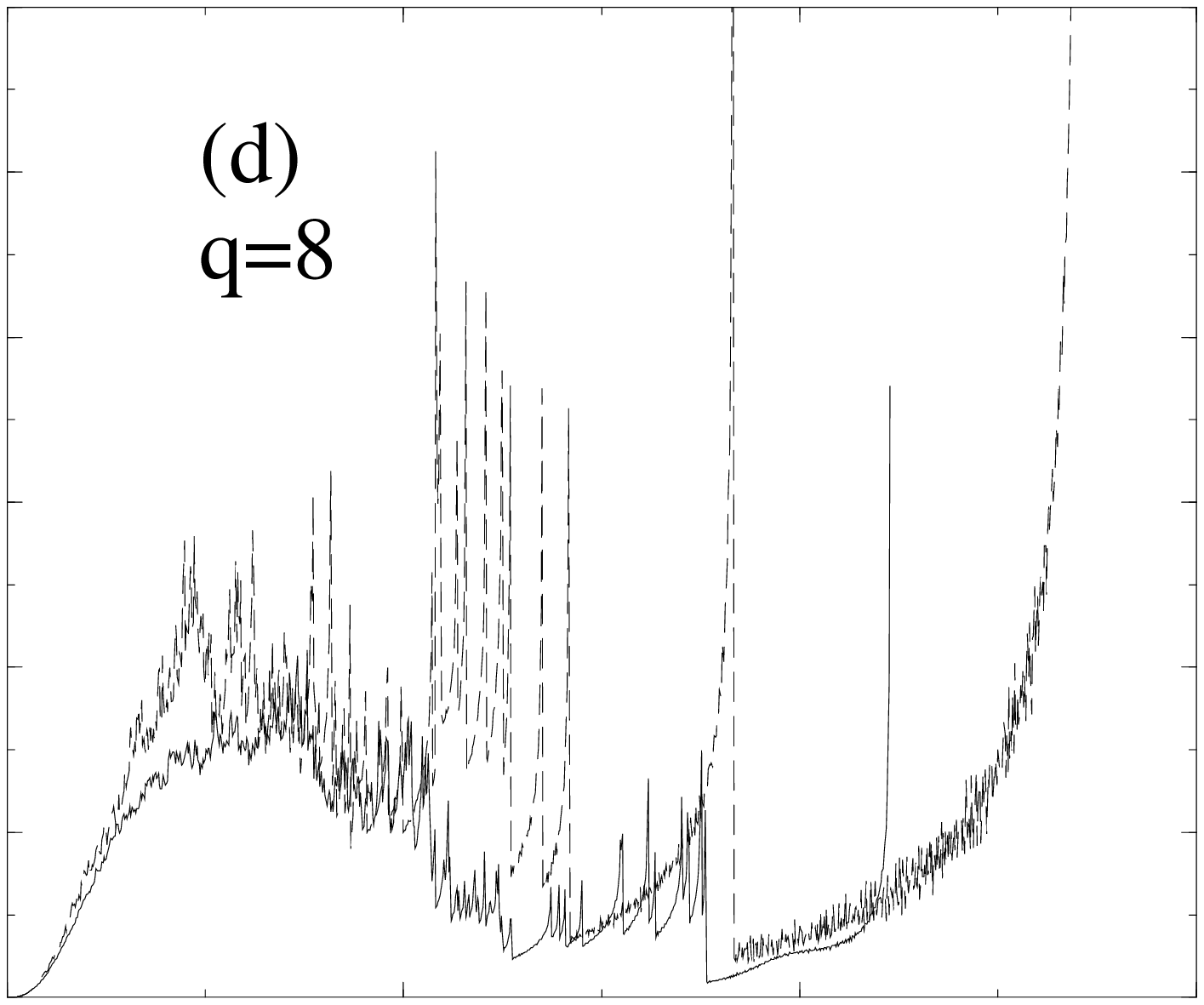}}

\bbox{\epsfxsize=4.3truecm\epsfbox{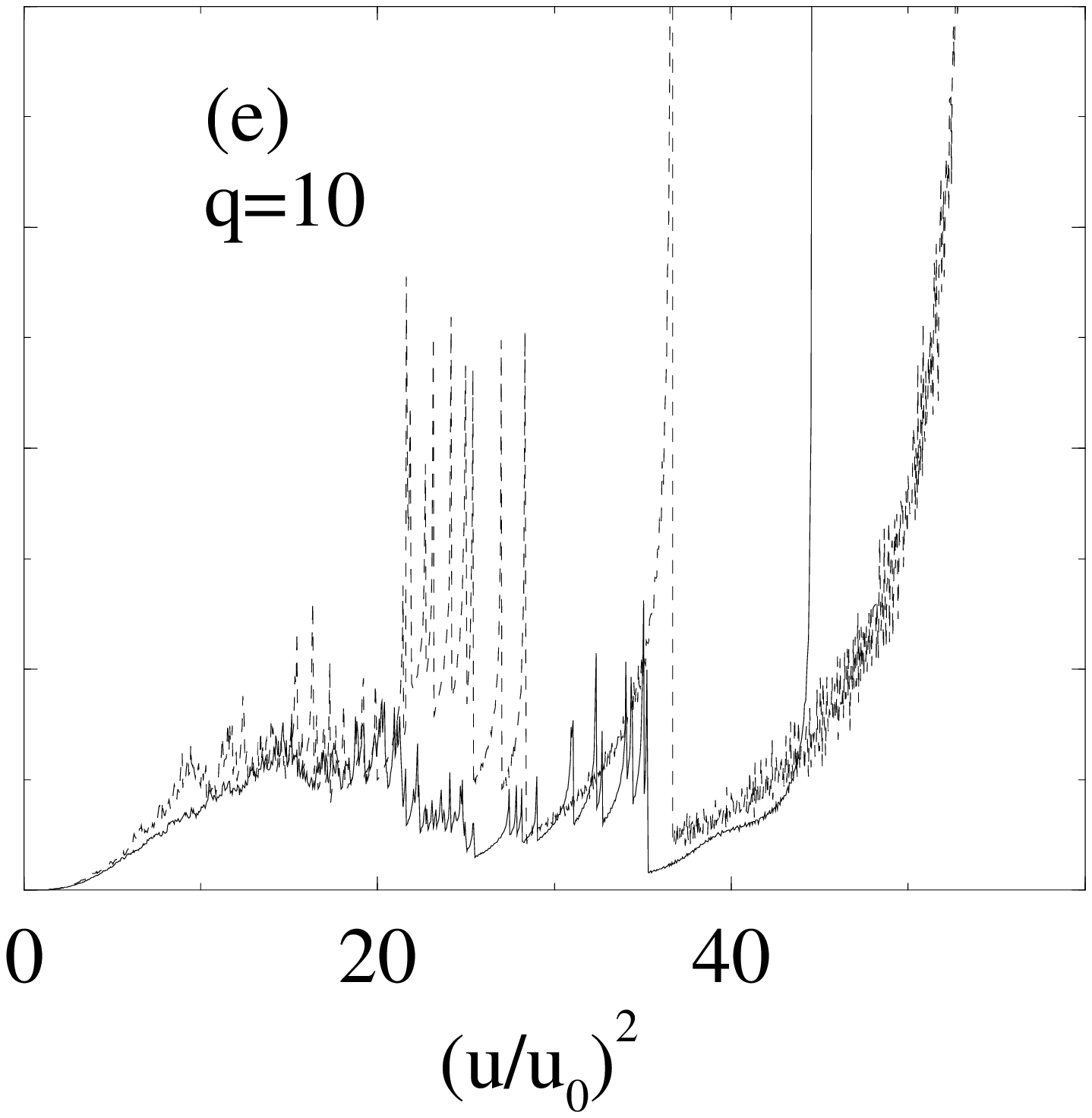}
\epsfxsize=4.3truecm
\epsfbox{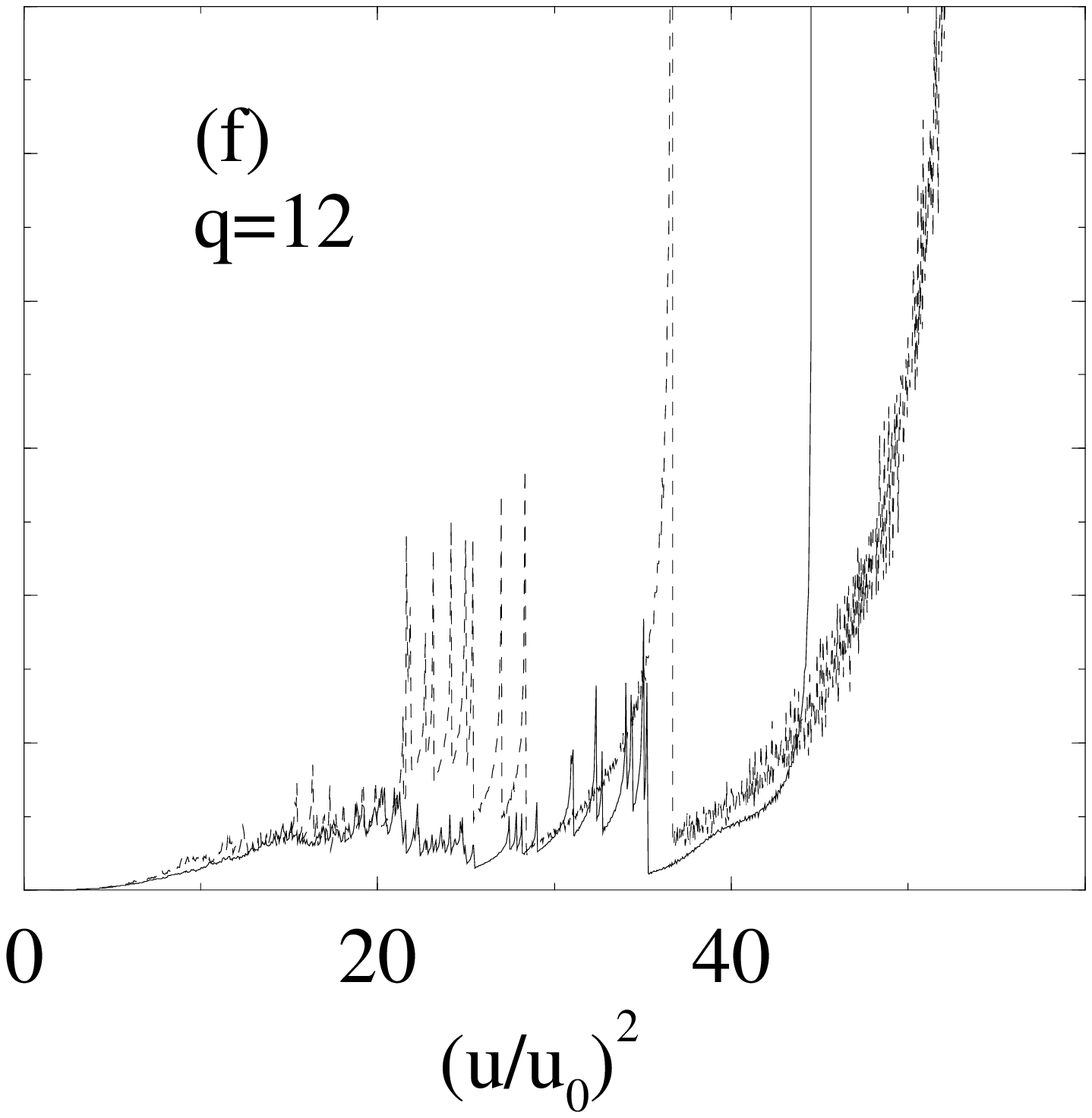}}
\caption{Same as figure 5  but for shell $\#7$. The solid line
represents a longer run of $4\times 10^4\tau_7$, and the dashed
line a shorter run of 4000$\tau_7$.}
\label{YY}
\end{figure}

\begin{figure}
\bbox{\epsfxsize=4.3truecm\epsfbox{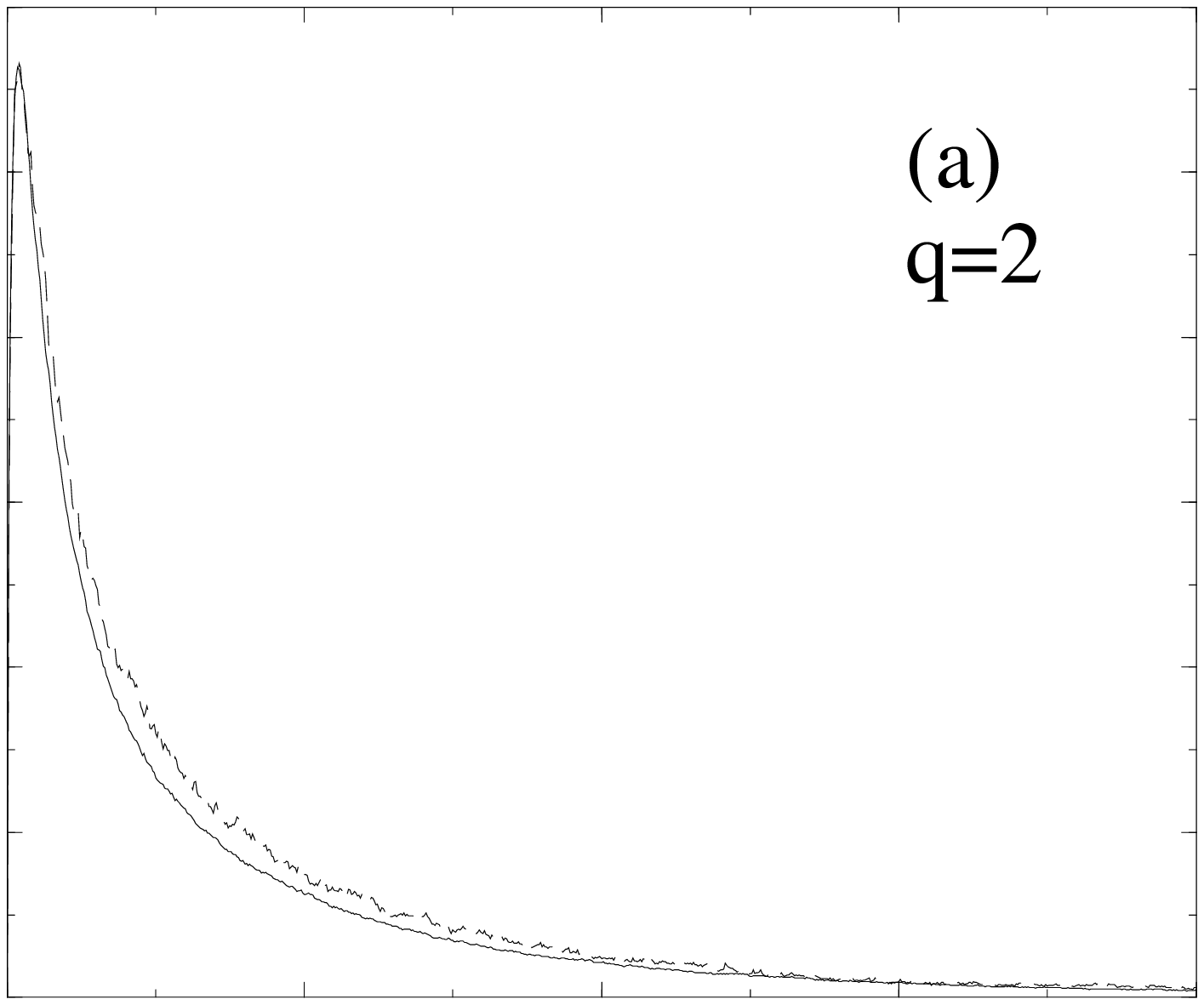}
\epsfxsize=4.3truecm\epsfbox{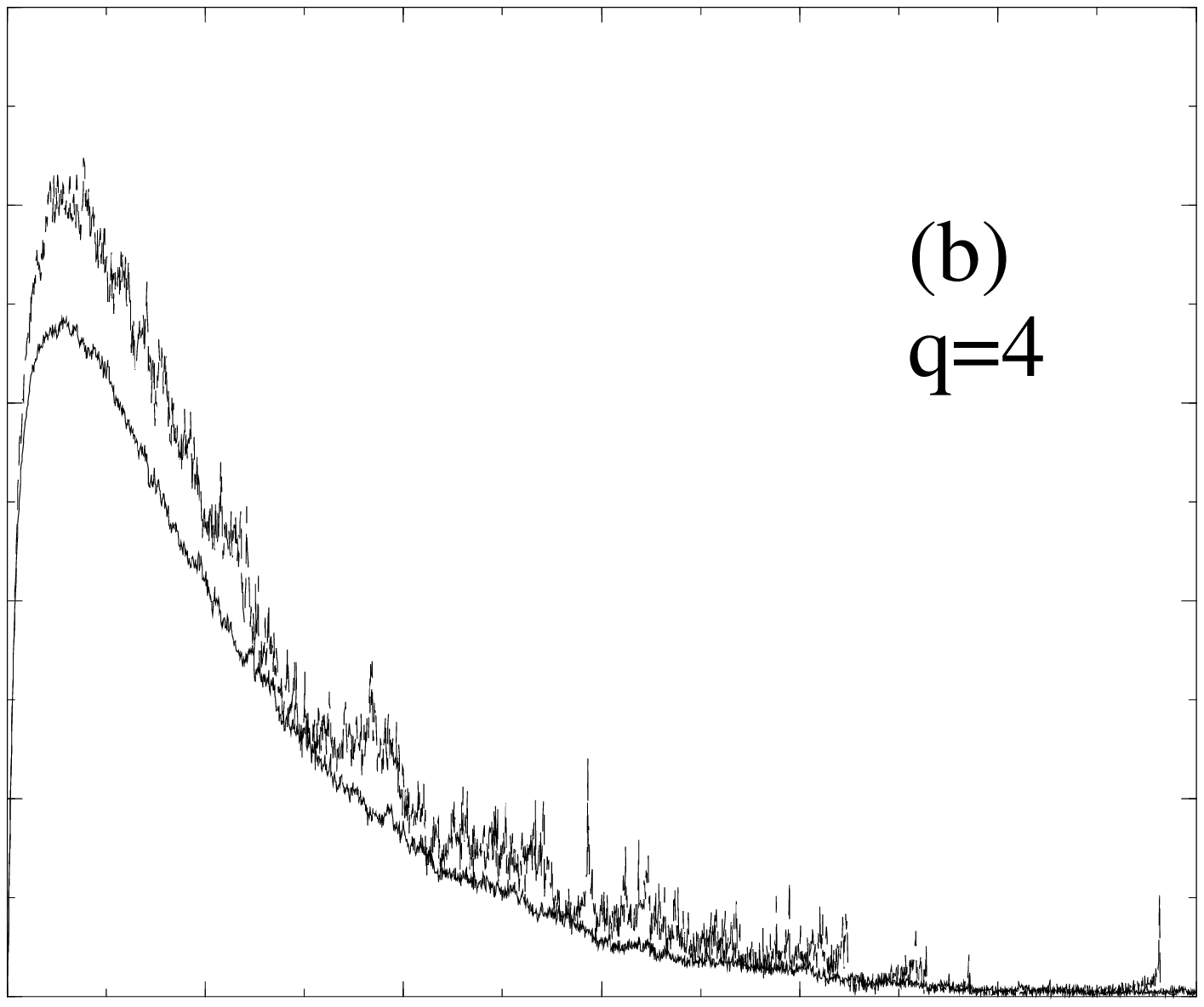} }

\bbox{\epsfxsize=4.3truecm\epsfbox{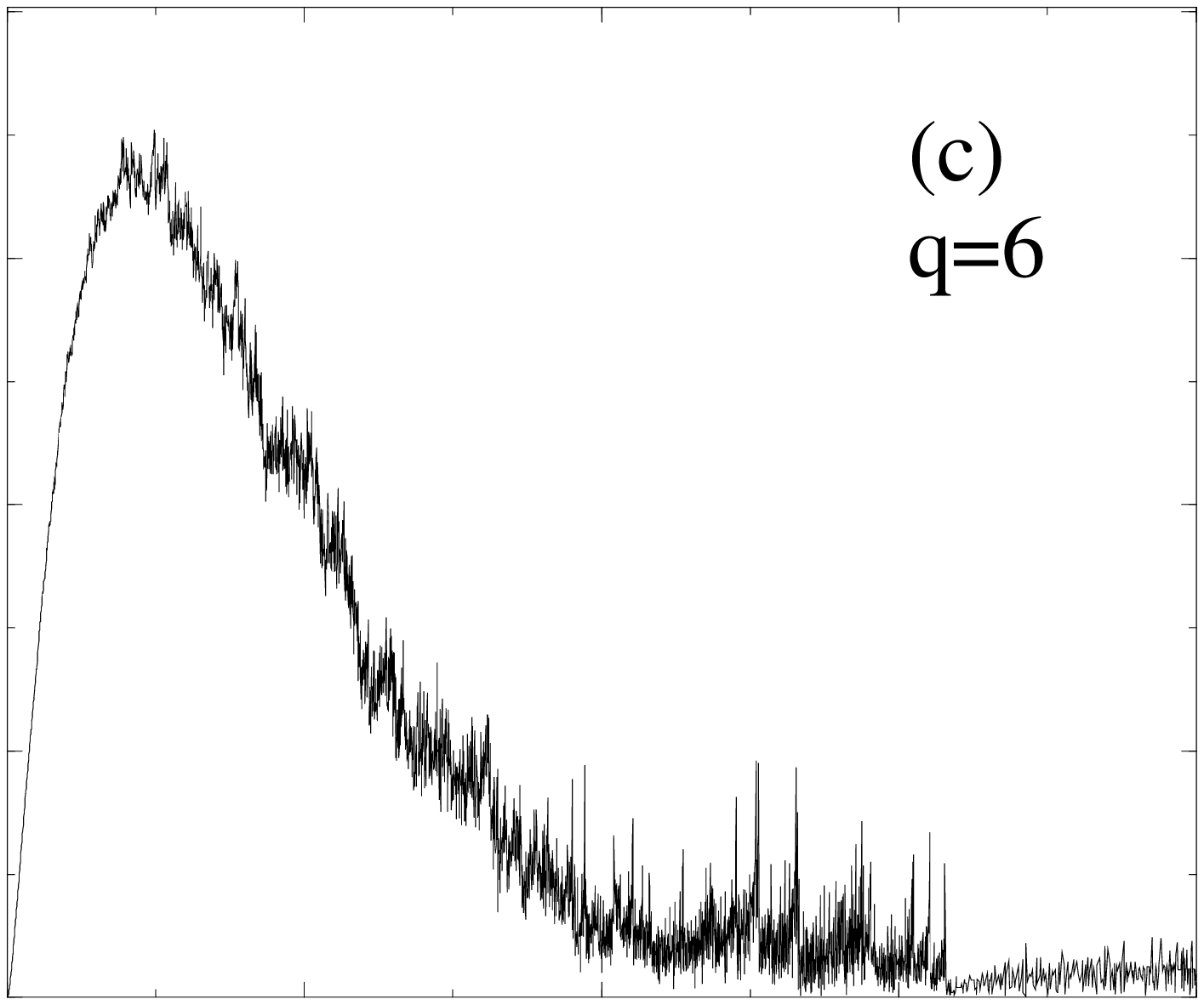}
  \epsfxsize=4.3truecm\epsfbox{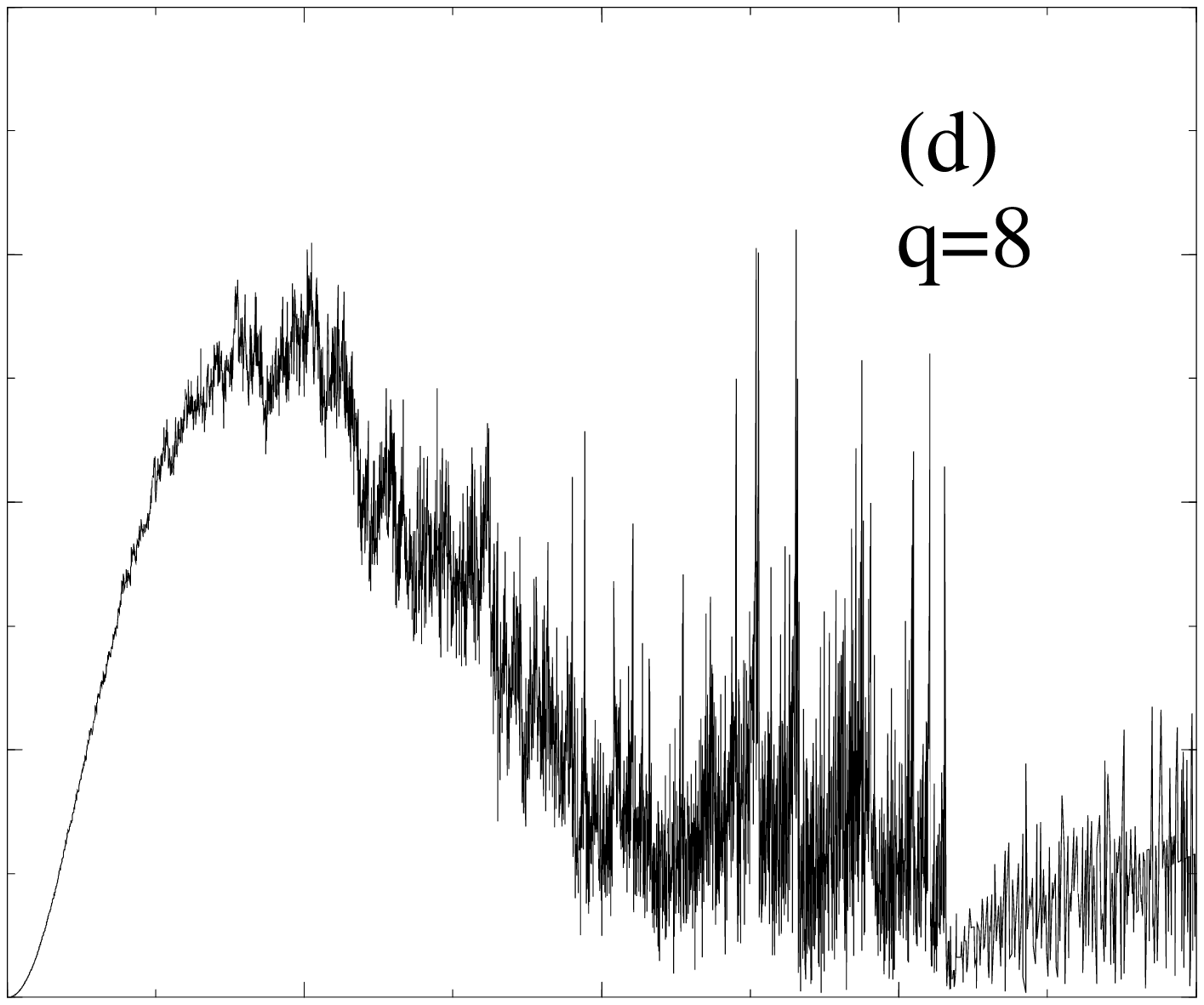} }

\bbox{\epsfxsize=4.3truecm \epsfbox{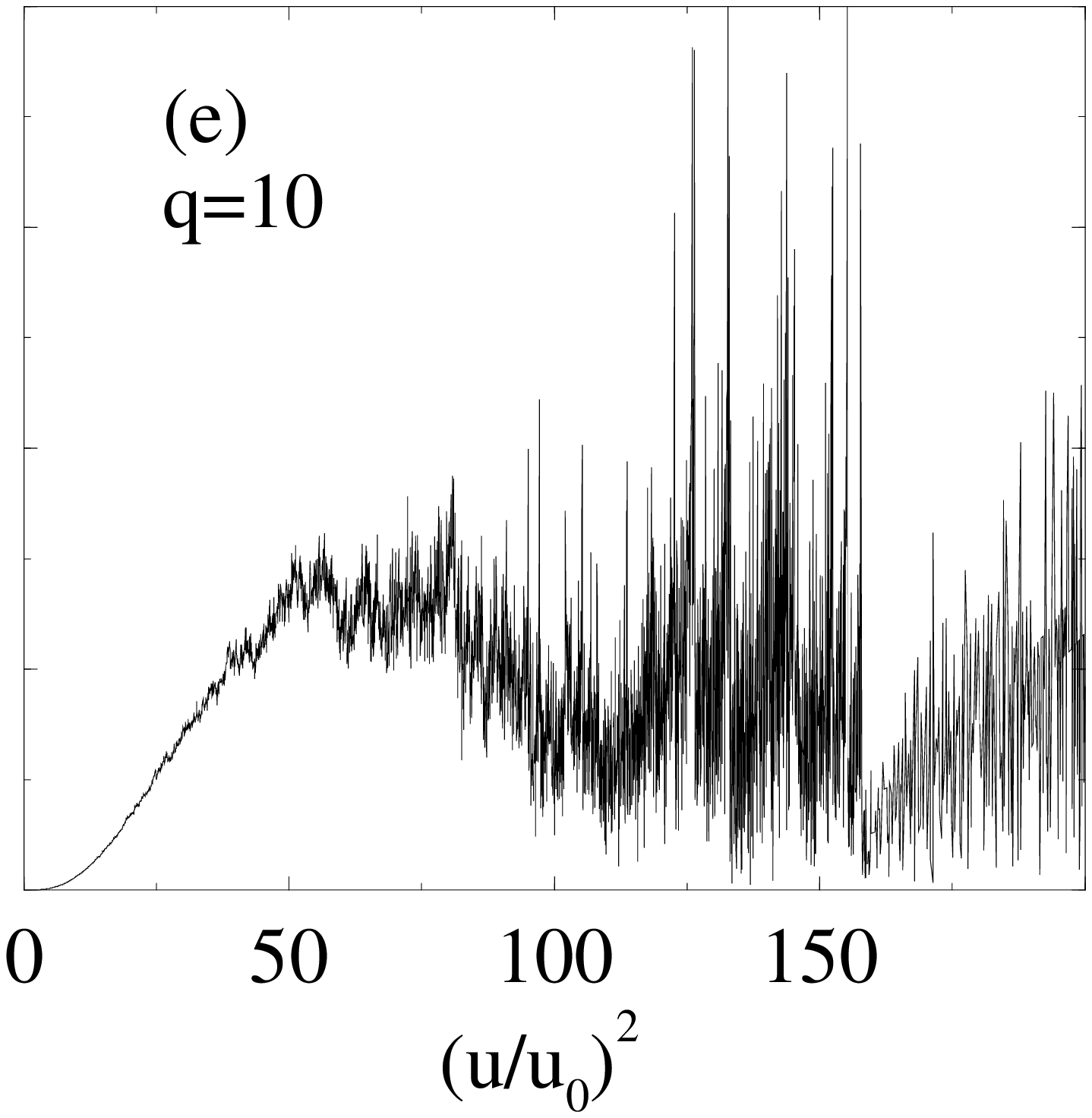}
\epsfxsize=4.3truecm \epsfbox{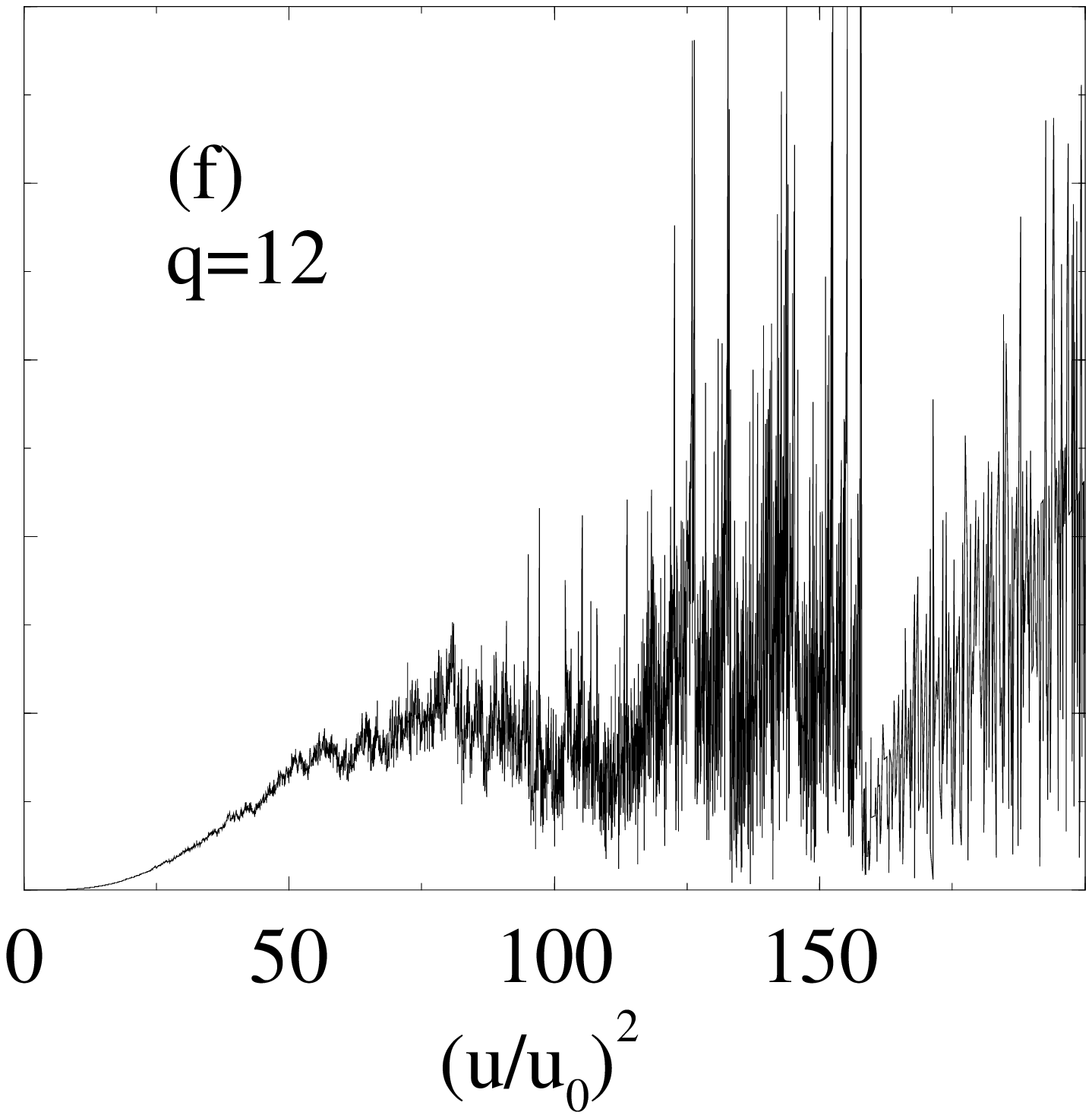}}

\caption{Same as figure 5  but for shell $\#12$. The solid line
represents a longer run of $4\times 10^5\tau_{12}$, and the dashed
line a shorter run of $4\times 10^4\tau_{12}$.}
\label{ZZ}
\end{figure}

\begin{figure}
\bbox{\epsfxsize=4.3truecm\epsfbox{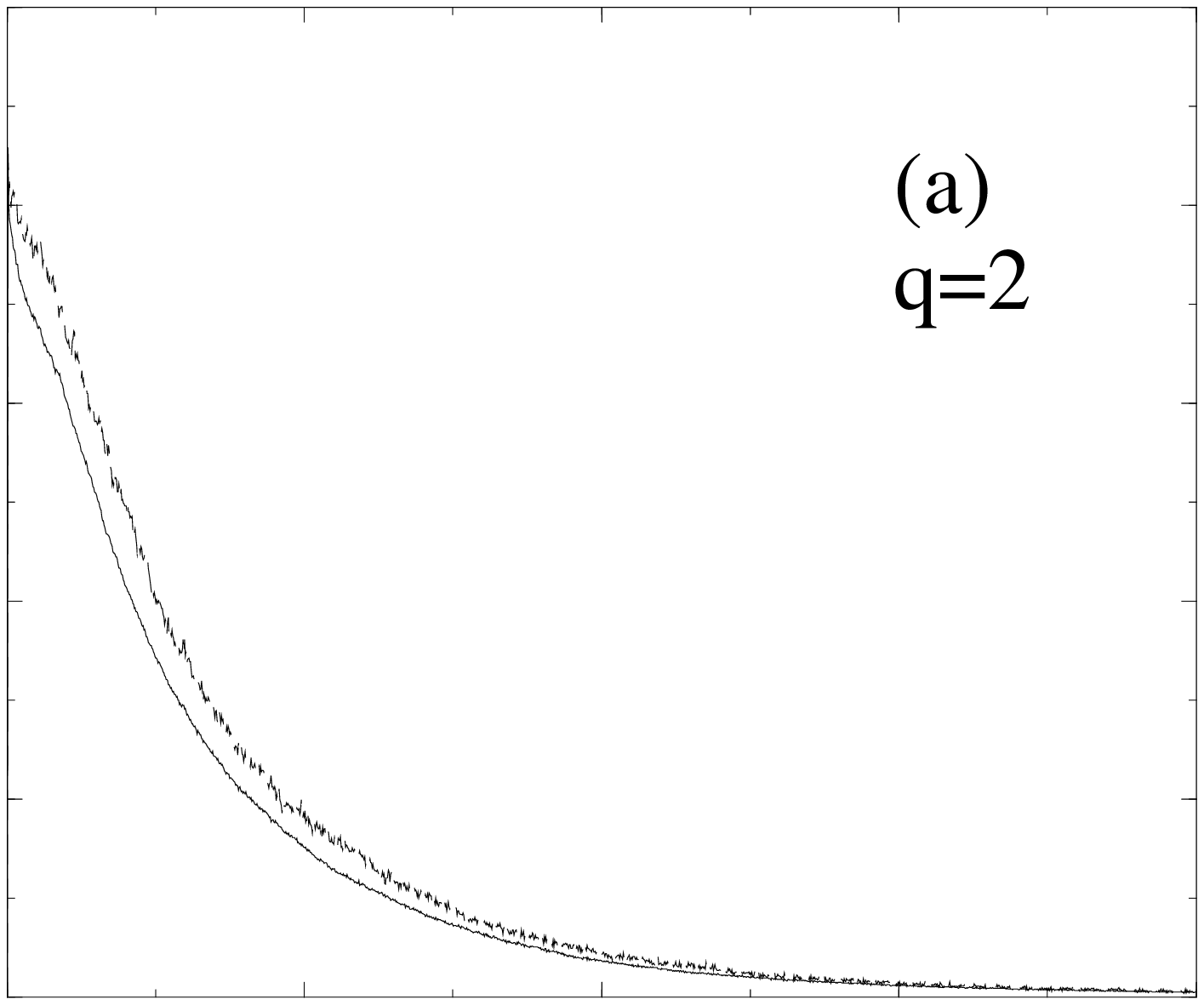}
\epsfxsize=4.3truecm\epsfbox{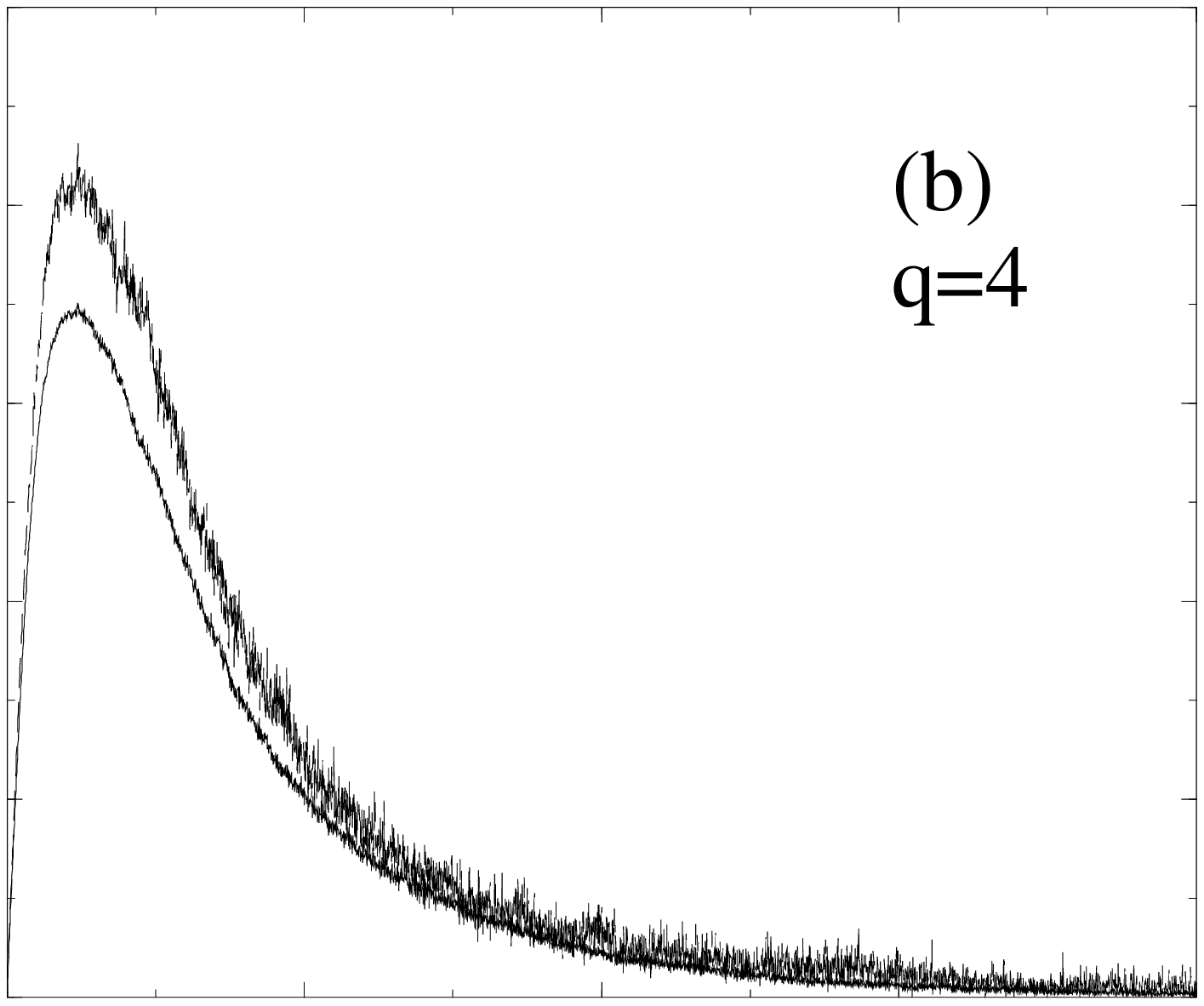}}
\bbox{\epsfxsize=4.3truecm\epsfbox{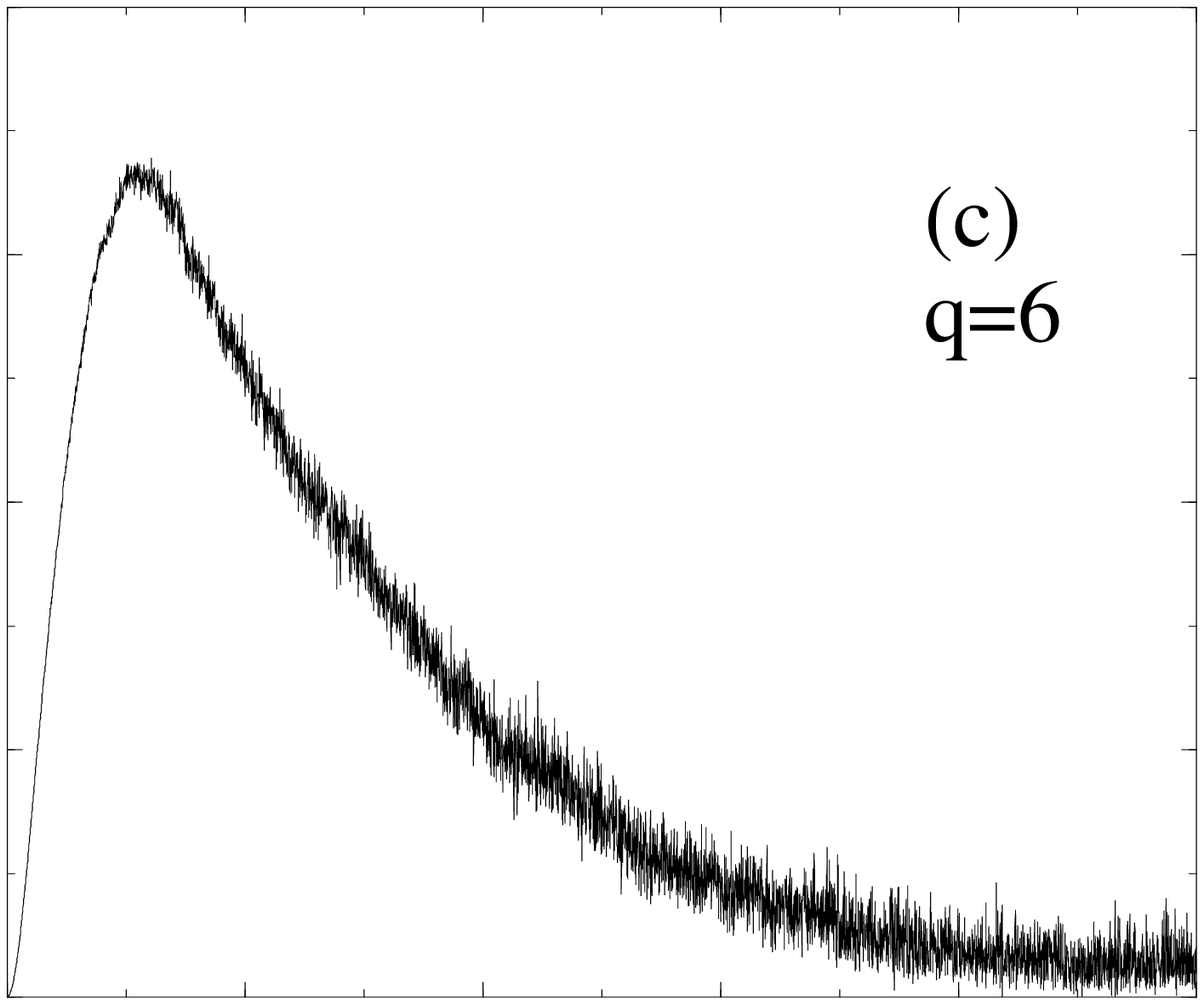}
\epsfxsize=4.3truecm\epsfbox{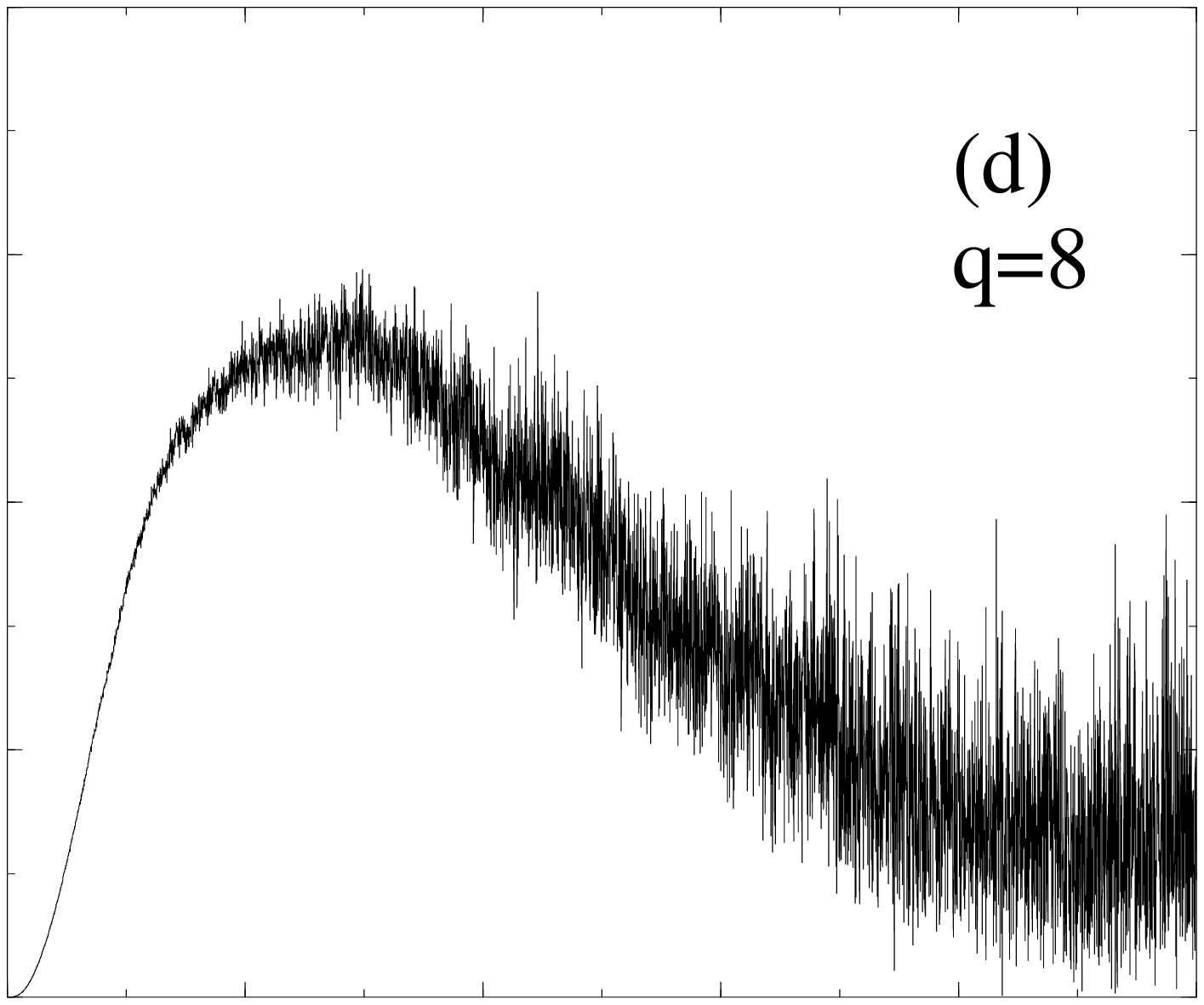}}
\bbox{\epsfxsize=4.3truecm\epsfbox{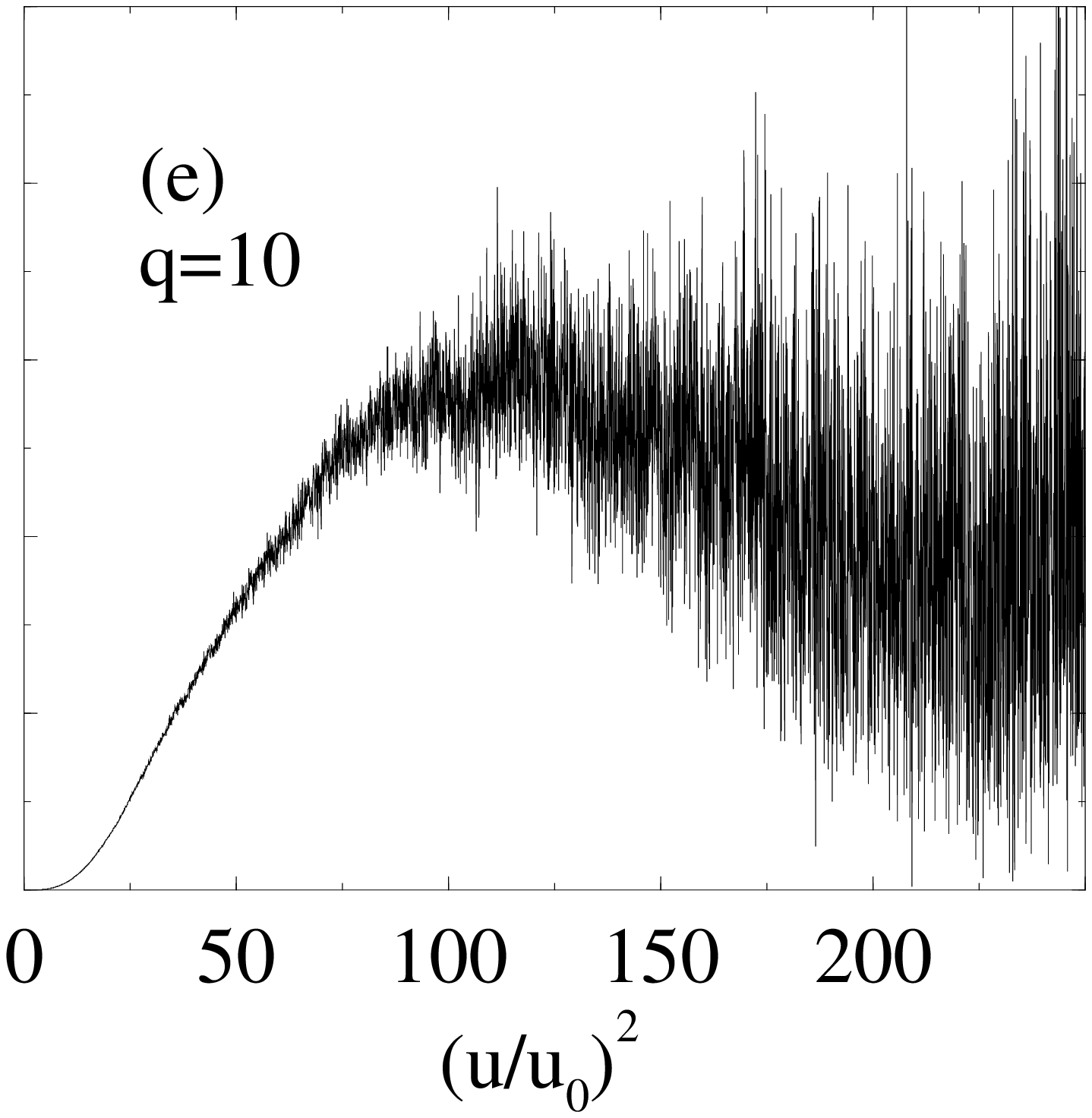}
\epsfxsize=4.3truecm\epsfbox{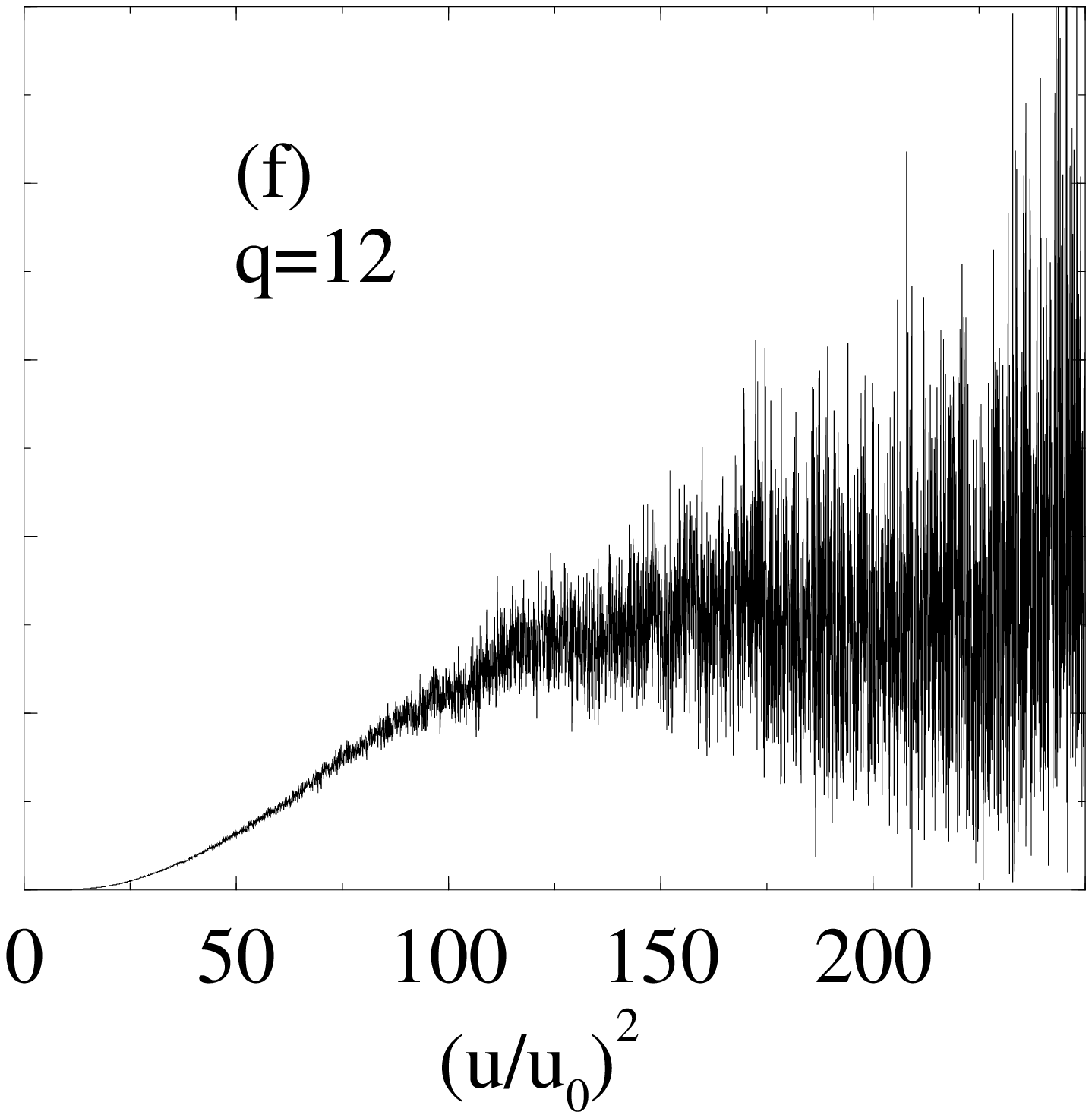}}
\caption{ Same as figure 5  but for shell $\#16$. The solid line
represents a longer run of $2.5\times 10^6\tau_{16}$, and the dashed
line a shorter run of $2.5\times 10^5\tau_{16}$. }
\label{WW}
\end{figure}
Admittedly this evaluation is rather rough. More accurate evaluations
should be based on the numerically computed probability distribution
functions as done for the GOY model in \cite{97LS}.  We plot
the numerical value of $(u/u_0)^{2q} P(u^2/u_0^2)$ versus $(u/u_0)^2$
and see how noisy is the region which gives the main contribution to
$S_{2q}$. Such plots for the third shell are presented at Fig~\ref{XX}
for two realizations, one averaged over 625 and the other over 6250
turnover times of this shell, $\tau_3$. In panels a, b, c we show the
integrands for $S_2$, $S_4$, and $S_6$.  One sees that $S_2$ and $S_4$
can be evaluated reasonably well even from the shorter run, while
$S_6$ can be computed only from the longer run. Panels d, e and f
present the analysis for $S_8$, $S_{10}$, and $S_{12}$
correspondingly.  The evaluation of $S_8$ is questionable even when
the long run is used; The results for $S_{10}$, $S_{12}$, etc.  are
meaningless even for the run of 6250 turnover-times. This run is too
short for this purpose.  The same analysis for shell \#7 (in the bulk
of the inertial interval), with two runs of $4000\tau_7$  
showing that the
improvement of the long run is not sufficient (see Fig. ~\ref{YY}).  We can hardly
compute $S_8$ from the longer run.  In the viscous end of the inertial
interval (say for the shell \#12) our runs was  ten times longer ($4\times
10^5 \tau_{12}$)  and the results can be seen in Fig.~\ref{ZZ}. 
 Now
$S_8$ can be computed reasonably well, but $S_{10}$ is still buried in
noise. Higher order structure functions cannot be estimated at all.
Lastly, in Fig.~\ref{WW} we present  results for the shell \#16 which
belongs to the beginning of the viscous subrange.  Here we have even
longer run of $2.5\times 10^6 \tau_{16}$, resulting in a marginal
improvement in the ability to compute $S_{10}$.

For the evaluation of the scaling exponent $\zeta_q$ one needs to
compute $S_q(k_n)$ throughout the inertial interval. It appears that
we can determine scaling exponents up to $\zeta_6$ from runs whose
duration is about 5000 (longest) turnover times.  In order to find
exponents up to $\zeta_8$ we need runs of minimal duration of $10^5$
(longest) turnover-times.  An  accurate determination of the exponent
$\zeta_{10}$ calls for runs of about one million turnover times!
Note that this estimate is in agreement with the simple analytical
formula (\ref{a7}) presented above.  Note also that these conclusions
may very well be applicable also for the analysis of experimental data
of hydrodynamic turbulence.  The scaling exponents with our choice of
parameters in the Sabra model correspond to those of Navier-Stokes
turbulence, and it is likely that the far end of the probability
distribution functions is as hard to reproduce in experiments as in
our simulations. Since very long runs are rarely available in
experimental data, this should serve as a warning that stated
numerical values of higher scaling exponents should be taken with
great caution.

\subsection{Test of the numerical procedure}
\label{62}
\begin{figure}
\epsfxsize=8.2truecm
\epsfbox{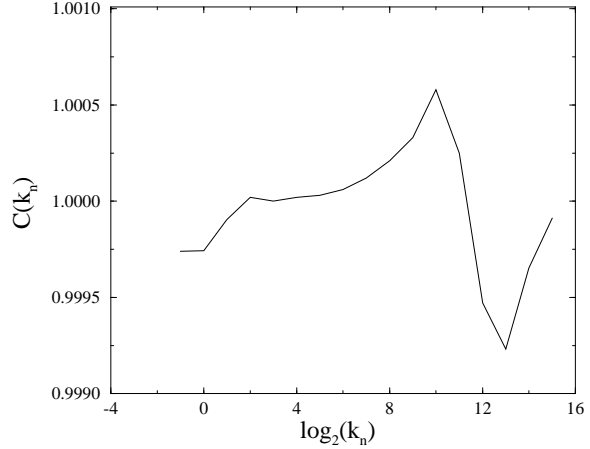}
\vskip 0.5cm 
\caption{Balance coefficient $C_n^{(2)}$ for 22 shells, $b=-0.5$,
  average over 2500 largest turnover times.}
\label{FigB}
\end{figure}

The averaging time is not the only factor affecting the quality of the
numerical data. Since the time dependence of $u_n(t)$ is highly
intermittent we need to test carefully the ability of the numerics to
cope with this. We need to check that the statistical characteristics
of the process $u_n(t)$ obey the exact relations imposed on the
correlation functions.  A simple test can be built around the first
equation of the infinite hierarchy relating $S_q (k_n)$ and $S_{q+1}(k_n)$.
Consider Eq.~(\ref{flux}) relating $S_2$ and $S_3$.  In the inertial
range, where the viscous term may be neglected, the largest term on
the LHS (proportional to $c$) is balanced by the two first terms on
the LHS.  In the viscous range, where $S_3(k_n)$ drops to zero very
quickly, this term is balanced by the viscous term on the RHS. It is
thus useful to rewrite Eq.~ (\ref{flux}) in the form of a "balance
coefficient " (keeping in mind that $S_3(k_n)$ is negative and
$S_2(k_n)$ is positive):

\begin{equation}   \label{A3}
C_n^{(2)}=  \frac { a |S_3(k_{n+1})|k_{n+1} + b |S_3(k_n)|k_n -
\nu k_n^{2} S_2(k_n) } {c |S_3(k_{n-1})|k_{n-1}}\  .
\end{equation}
If the numerical data satisfies the balance equation (\ref{flux})
accurately, the coefficients $C_n^{(2)}$ has to be unity for all $n$.
In Fig.~\ref{FigB} we show that in our simulations this relation
between $S_3(k_n)$ and $S_2(k_n)$ is obeyed with accuracy better than
0.1\%.  However, this does not mean that less frequent events which
contribute to higher order correlation functions are also correctly
reproduced. To check the statistical reliability of $S_4(k_n)$ one can
use the second equation from the hierarchy, which connects $S_4(k_n)$
and $S_5(k_n)$ and so on. To measure this accuracy one can define,
analogously to $C_n^{(2)}$, a generalized balance coefficient $C_n^{(2q)}$. To
define it we consider the time derivative of $S_{2q}(k_n)$:
\begin{eqnarray}\nonumber
&&{dS_{2q}(k_n)\over dt}= -2q {\rm Im }\Big[ak_{n+1}
\left<u_n^*u_{n+1}^*u_{n+2}|u _n|^{2(q-1)}\right> \\  && 
+bk_n\left<u_{n-1}^*u_n^*u_{n+1}|u
  _n|^{2(q-1)}\right> \label{a8}   \\
&&  -ck_{n-1}\left<u_{n-2}u_{n-1}u_n^*|u_n|^{2(q-1)}\right>\Big] -2q\nu
k_n^{2}\left<|u_n|^{2q}\right>. \nonumber \end{eqnarray}
In the stationary case this gives:
\begin{eqnarray} \nonumber
&&a S_{2q+1}^-(k_{n+1}) k_{n+1} +b S_{2q+1}(k_n) k_n \\ \label{a9} &&
+c S_{2q+1}^+(k_{n-1}) k_{n-1}= \nu k_n^{2} S_{2q}(k_n) .
\end{eqnarray}
Here $S_{2q+1}$ is defined by Eq.(\ref{Ssabra}) and we have introduced two
additional structure functions: 
\begin{equation}\label{a10}
S_{2q+1}^\pm(k_{n})={\rm Im}\left< u_{n-1}u_nu_{n+1}^*
|u_{n\pm1}|^{2(q-1)}\right>.
\end{equation}
One can rewrite (\ref{a9}), similarly to (\ref{A3}), in the form of a  
balance coefficient:
\begin{eqnarray}\label{a11}
&& C_n^{(2q)} \\ \nonumber
&=&  \frac { a S_{2q+1}^-(k_{n+1})k_{n+1} 
+ b S_{2q+1}(k_n) k_n + \nu k_n^{2} 
S_{2q}(k_n) } {c |S_{2q+1}^+(k_{n-1})|k_{n-1}} .
\end{eqnarray}
Again, if the numerical data reproduce the balance equation
(\ref{a9}), the coefficient $C_n^{(2q)}$ has to be unity for
all $n$. Testing this fact should be an integral part of the
numerical solution of this model and similar models in the future.
\begin{figure}
\epsfxsize=8.2truecm
\epsfbox{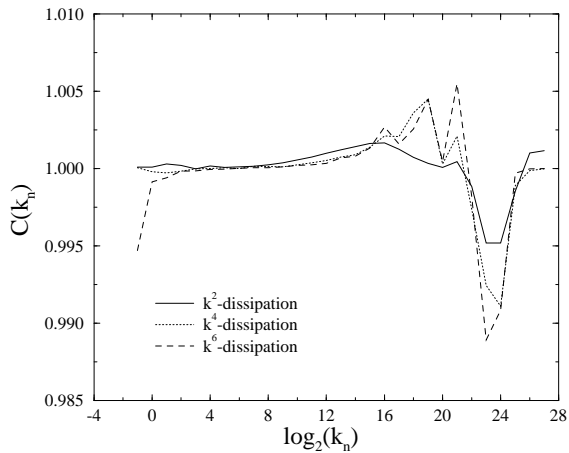}
\caption
{Coefficients $C_n^{(2)}$ [see Eq. (55) ] obtained 
for a model of 34 shells integrated over 250 forcing turnover time
scales with dissipative terms proportional to $k_n^2$, $k_n^4$ and
$k_n^6$ respectively.}
\label{sabrabalance34}
\end{figure}
\section{Universality with respect to Hyperviscosity }
``Hyperviscosity" in shells models amounts to changing the viscous
term in (\ref{sabra}) with a term $\nu k_n^{2m}$ with $m>1$. The
effect of hyperviscosity on shell models is a matter of controversy.
It has originally been argued by She and Leveque \cite{SL95PRL} that
in the GOY model there was no universality of the scaling exponents,
the value of the latter being strongly dependent on the dissipation
mechanism. The same observation has been made by
Sch{\"o}rghofer\cite{Schorg95}  {\it et al} and by Ditlevsen
\cite{DitPF97}.  If true this observation would cast a doubt either on
the relevance of shell models in turbulence studies or on one of the
most widely accepted hypotheses in fluid turbulence: the universality
of the exponents in the scaling range. Note for example that many
direct simulation of 3D turbulence use hyperviscosity.  On the other hand,
Benzi {\it et al} have showed  \cite{98BBST} that in
the case of shell models with eddy viscosity, the inertial exponents
were independent on the particular definition used for the  eddy
viscosity. We made
recently the point \cite{98LPV} that within the GOY model this
phenomenon is nothing but a finite size effect which disappears when
one increases the size of the inertial range. We dedicate this Section
to showing that the same is true for the Sabra model.

Before discussing the results we need to test our simulations for
accuracy of the evaluation of the structure functions. To this aim we
present in Fig.~\ref{sabrabalance34} the balance coefficient $C_n^{(2)}$
(cf. Sect \ref{62}) for $m=1,2,3$. We tested the accuracy in a
relatively short run of 250 forcing turnover times scales. The results
indicate that even for this short run the accuracy of determination of
the two lowest order structure functions is about 0.1\% in the
inertial range, but only about 1\% in the dissipative range. Note that
hyperviscosity makes the determination of the structure functions in
the viscous range (starting with the crossover region) somewhat less
accurate. In order to
reduce the source of uncertainty and without loss of generality, we
measured the exponents from the flux based structure functions (\ref{hatSp}).

In the following, we focus on the second and third order structure
functions,  using runs of
duration 1500 forcing turnover times scales, and offer a
careful calculation of their apparent scaling exponents as a function
of the number of shells used in the simulations, and of the order of
the hyperviscosity term $m$.  We will show that the hyperviscous
correction affects a finite number of shells in the vicinity of the
viscous transition. This number is relatively large, about ten shells
or three decades of ``length-scales". The reason for this large effect
is that we have a discrete model in which each shell interacts with 4
nearest neighbors. This means that with the standard shell spacing
parameter $\lambda=2$, the local interactions spread over more than one
decade of length scales. Nevertheless, we show now that this number
remains unchanged when we increase the size of the inertial range,
indicating a mere finite size effect.

\begin{figure}
\epsfxsize=8.2truecm
\epsfbox{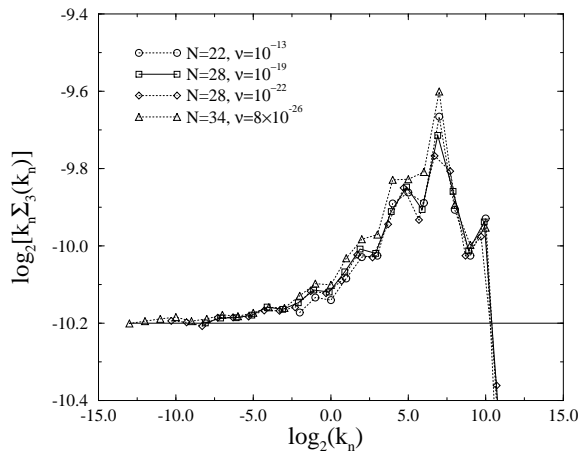}
\caption
{Log-log plots of $k_n\Sigma_3(k_n)$ vs. $k_n$ in case of hyperviscosity of
index $m=2$ with different numbers of shells and viscosities.  The
collapse has been obtained by shifting the abscissa. The solid line
shows the constant behavior expected theoretically. One observes
clearly that the departure from this constant value only occurs in a
region of about ten shells near to the viscous transition. When the
inertial range is large enough, the predicted behavior is recovered.}
\label{k4collapse}
\end{figure}

\begin{figure}
\epsfxsize=8.2truecm
\epsfbox{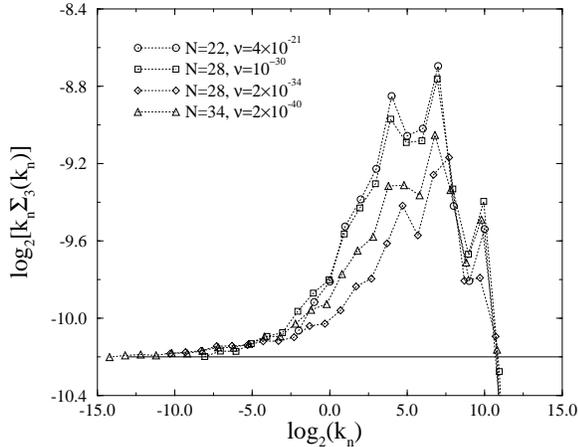}
\caption
{Same figure as above in the case of hyperviscosity of index
$m=3$. Note that the amplitude of the bump is larger than in the
previous case.}
\label{k6collapse}
\end{figure}

To see this point examine Figs. \ref{k4collapse} and \ref{k6collapse}
in which we superpose results for $k_n\Sigma_3(k_n)$ with $m=2$ and $m=3$
respectively, which were obtained in eight different simulations as
detailed in the figures.  The plots are as a function of $\log(k_n)$
with an appropriate shift in the abscissa. We see that in all cases the
region of deviation from a constant function, associated with the
theoretical expectation Eq.~(\ref{scaling}) is of constant magnitude
and of constant extent, independent of $\nu$ or the total number of
shells. This is a clear indication that when the number of shells
increases to infinity the scaling exponent $\zeta_3=1$ will be
observed in a universal manner.

\begin{figure}
\epsfxsize=8.2truecm
\epsfbox{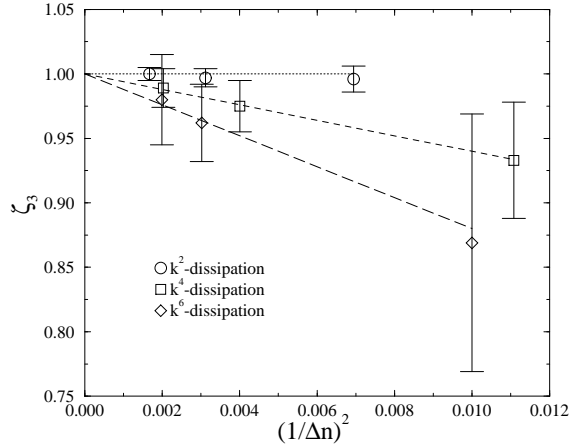}
\caption{The apparent scaling exponent $\zeta_3$ as a function of the
  square of the inverse extent of the inertial interval, for for $m=1$
  (circles), $m=2$ (squares) and $m=3$ (diamonds). The tendency
  towards $\zeta_3=1$ is evident.}
\label{S3k4k6}
\end{figure}

\begin{figure}
\epsfxsize=8.2truecm
\epsfbox{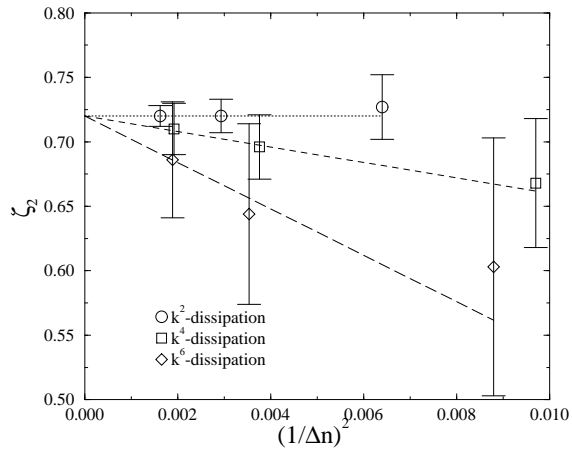}
\caption{The apparent scaling exponent $\zeta_2$ as a function of the
  square of the inverse extent of the inertial interval, for $m=1$
  (circles), $m=2$ (squares) and $m=3$ (diamonds). The tendency
  towards $\zeta_2$ as found for normal viscosity $m=1$ is evident.}
\label{S2k4k6}
\end{figure}

Another way to reach the same conclusion is obtained by fitting
structure functions as explained in section 6 to the formula
(\ref{fit}). We ran simulations for $m=1,2$ and 3 with $N=22,28$ and
34.  The exponent $x$ of the viscous tail for $m=2,3$ exhibits
significant departures from its dimensional expectation (\ref{def-x}). We
obtained $x\simeq 0.75$ for $m=2$ and $x\simeq 0.90$ for $m=3$, while
$x\simeq 0.69$ for $m=1$. These values which seem to be independent on
the order of the structure function have then been used in the fitting
procedure.  The results for $\zeta_2$ and $\zeta_3$ with normal
viscosity were quite independent of $N$. On the contrary,
hyperviscosity caused an apparent change in scaling exponents.
However, as can be seen in Figs.  \ref{S3k4k6} and \ref{S2k4k6} these
values can be plotted as a function of $1/[\log (k_d/k_1)]^2$ and they
converge, for $k_d\to \infty$ to the values obtained for $m=1$. Note
that $\log(k_d/k_1)$ is precisely the length of the inertial interval,
and $k_d$ was obtained from the fit.
\section{Summary}
We presented a new shell model of turbulence, and demonstrated its
improved properties in terms of simpler, shorter range correlations.
The model exhibits anomalous scaling similarly to the GOY model and
to Navier-Stokes turbulence. In the future we will argue that the improved
properties of this model help considerable in seeking analytic methods
for the calculations of the scaling exponents.
We used the opportunity of the introduction of this model to examine
carefully issues like the accuracy of determination of scaling exponents
and the minimal length of running time required to achieve accurate
structure functions. These considerations are model independent and
pertinent to other examples of multiscaling as well.
Lastly, we demonstrated the universality of the scaling exponents with
respect to the type of viscous damping. This universality was questioned
in the recent literature but we showed here for the Sabra model and 
previously \cite{98LPV} for the GOY model that there is no reason to doubt
it.

\vspace{10pt}
\noindent{\large \bf Acknowledgments}

This work was supported in part by the US-Israel Bi-National Science
Foundation, The Basic Research Fund administered by the Israel Academy
of Science and Humanities, The European Union under contract
FMRX-CT-96-0010 and the Naftali and Anna Backenroth-Bronicki Fund for
Research in Chaos and Complexity.


\begin{thebibliography}{99}

\bibitem{Gledzer}
E.~B. Gledzer.
 Dokl. Akad. Nauk. SSSR  {\bf 200}, 1043, (1973).

\bibitem{GOY}
M.~Yamada and K.~Ohkitani.
 J. Phys. Soc. Jpn. , {\bf 56}, 4210, (1987).

\bibitem{Jensen91PRA}
M.~H. Jensen, G.~Paladin, and A.~Vulpiani.
 Phys. Rev. A  {\bf43(2)}, 798, (1991).

\bibitem{Piss93PFA}
D.~Pissarenko, L.~Biferale, D.~Courvoisier, U.~Frisch, and M.~Vergassola.
 Phys. Fluids A, {\bf 5(10)}, 2533, (1993).

\bibitem{Benzi93PHD}
R.~Benzi, L.~Biferale, and G.~Parisi.
 Physica D {\bf  65}, 163, (1993).

\bibitem{98BLPP}
V.I. Belinicher, V.S. LÕvov, A. Pomyalov, and I. Procaccia, 
ÒComputing the Scaling  Exponents in Fluid Turbulence from First Principles: 
Demonstration of Multi-scalingÓ,        J. Stat. Phys.,  submitted.

\bibitem{SLATEC} Slatec library (Sandia, Los Alamos, Air Force
Weapons Laboratory Technical Exchange Committee) available on
http://www.netlib.org/slatec

\bibitem{DDEBDF} L.F.~ Shampine and H.A. Watts SAND-79-2374, DEPAC 

\bibitem{Fox} R.F. Fox, I.R. Gatland, R. Roy and G. Vemuri 
Phys. Rev. A  {\bf 38}, 5938 (1988)

\bibitem{SL95PRL}
E.~Leveque and Z.S. ~She.
Phys. Rev. Lett. , {\bf 75(14)}, 2690, (1995).
\bibitem{97LS}  E.~Leveque and Z. S. ~ She Phys. Rev. E {\bf 55} 2789 (1997).
\bibitem{93Ben}
R.~Benzi, S.~ Ciliberto, R. ~Tripiccione, C. ~Baudet, F. ~Masaioli and 
S.~ Succi, Phys. Rev. E {\bf 48}, R29 (1993).


\bibitem{Schorg95}
N.~Sch{\"o}rghofer, L.~Kadanoff, and D.~Lohse.
Physica D, {\bf 88}, 44--54 (1995).

\bibitem{DitPF97}
P.~Ditlevsen.
Phys. Fluids {\bf 9(5)} 1482, 1997.

\bibitem{98LPV}
V.S. L'vov, I. Procaccia and D. Vandembroucq.
{\sl Universal scaling exponents in shell models of turbulence: 
Viscous effects are finite-size corrections to scaling}. 
Phys. Rev. Lett, submitted. Chao-dyn \# 9803014
\bibitem{98BBST}
R. Benzi, L. Biferale, S. Succi, F. Toschi.
{\sl Intermittency and eddy-viscosities in dynamical models of
turbulence}.
Chao-Dyn 9802020, submitted to Physics of Fluids


\end{thebibliography}
\end{document}